\documentclass[12pt]{iopart}
\usepackage{graphicx}
\usepackage{bm}
\usepackage{IEEEtrantools}
\usepackage{amsthm}
\usepackage{algorithm}
\usepackage{algpseudocode}
\usepackage{siunitx}
\usepackage{caption}
\usepackage{subcaption}
\usepackage{tabularx}
\usepackage{float}
\usepackage{iopams}

\newtheorem{proposition}{Proposition}

\begin{document}

\title[Label-free timing for modularized detectors]{Label-free timing analysis of SiPM-based modularized detectors with physics-constrained deep learning}

\author{Pengcheng Ai$^{1,2,*}$, Le Xiao$^{1,2,*}$, Zhi Deng$^{3}$, Yi Wang$^{3}$, Xiangming Sun$^{1,2}$, Guangming Huang$^{1,2}$, Dong Wang$^{1,2}$, Yulei Li$^{3}$ and Xinchi Ran$^{3}$}

\address{$^1$ PLAC, Key Laboratory of Quark and Lepton Physics (MOE), Central China Normal University, Wuhan, 430079, China}
\address{$^2$ Hubei Provincial Engineering Research Center of Silicon Pixel Chip \& Detection Technology, Wuhan, 430079, China}
\address{$^3$ Key Laboratory of Particle and Radiation Imaging (MOE), Department of Engineering Physics, Tsinghua University, Beijing, 100084, China}
\address{$^*$ Authors to whom any correspondence should be addressed.}

\ead{aipc@ccnu.edu.cn and lxiao@mail.ccnu.edu.cn}

\vspace{10pt}
\begin{indented}
\item[]\today
\end{indented}

\begin{abstract}
Pulse timing is an important topic in nuclear instrumentation, with far-reaching applications from high energy physics to radiation imaging. While high-speed analog-to-digital converters become more and more developed and accessible, their potential uses and merits in nuclear detector signal processing are still uncertain, partially due to associated timing algorithms which are not fully understood and utilized. In this paper, we propose a novel method based on deep learning for timing analysis of modularized detectors without explicit needs of labelling event data. By taking advantage of the intrinsic time correlations, a label-free loss function with a specially designed regularizer is formed to supervise the training of neural networks towards a meaningful and accurate mapping function. We mathematically demonstrate the existence of the optimal function desired by the method, and give a systematic algorithm for training and calibration of the model. The proposed method is validated on two experimental datasets based on silicon photomultipliers (SiPM) as main transducers. In the toy experiment, the neural network model achieves the single-channel time resolution of 8.8 \si{ps} and exhibits robustness against concept drift in the dataset. In the electromagnetic calorimeter experiment, several neural network models (FC, CNN and LSTM) are tested to show their conformance to the underlying physical constraint and to judge their performance against traditional methods. In total, the proposed method works well in either ideal or noisy experimental condition and recovers the time information from waveform samples successfully and precisely.
\end{abstract}

%
\vspace{2pc}
\maketitle
%
%

\section{Introduction}

Pulse timing is an important research topic in nuclear detector signal processing, with applications ranging from high energy physics to radiation imaging. Accurate time measurements are meaningful for precisely determining the vertex of interactions as well as the dynamics of incident particles. In the past decade, high-speed analog-to-digital converters (ADC) were designed for front-end electronics of nuclear detectors \cite{AMELI2019286} and incorporated into their electronic dataflow. Traditionally, some fixed algorithms can be used to extract time information from a time series of pulse samples, such as leading edge discrimination or constant fraction discrimination \cite{FALLULABRUYERE2007247}. However, when they work in noisy or changing conditions, the performance of these fixed algorithms drops significantly \cite{Ai_2021}.

On the other hand, machine learning techniques, especially neural networks (NN) and deep learning, open another door for possible solution of time extraction from waveform samples. Recent literature has demonstrated that NNs can approximate the Cram\'er Rao lower bound in a broad range of working conditions \cite{Ai_2021}. It is estimated that NNs, the key components of intelligent front-end processing, will be widely used in future nuclear detector systems \cite{Humble_2022,Braga2022,Therrien:22,Carini2022}, empowered by ever-growing developments of hardware acceleration for NN inference \cite{Aarrestad_2021,Khoda_2023,Ngadiuba_2021}.

However, one pre-requisite for NNs to achieve superior performance is a justified reference (ground-truth label) in training. While labelled data are easily available in computer vision and many other machine learning tasks, they are not so for nuclear detector experiments. To provide accurate time references for real-world nuclear detectors, additional timing instruments or specific experimental schemes are needed. For example, in Ref. \cite{GLADEN2020164505}, the timing resolution of a high purity germanium detector was optimized based on the timing reference of a $\mathrm{BaF}_2$ scintillation detector. In Ref. \cite{Carra_2022}, the timing resolution and depth-of-interaction of monolithic scintillators were studied with a collimator crystal placed on the opposite side of the radioactive source. In Refs. \cite{Berg_2018,Kwon2021,10038575}, a back-to-back experimental setup was established to capture coincidence events, while NNs were used to predict the timing difference of paired detectors. In Ref. \cite{Onishi_2022}, NNs generated continuous value predictions based on the timing reference of leading edge discrimination. Although impressive results have been obtained in these studies, the significance of machine learning is limited either by the procedure (external calibration equipment) or the predicted target (preset timing differences).

Therefore, it is worthwhile to exploit the built-in structure of nuclear detectors for potential timing correlations and make use of them in the training process. The idea of training NNs without explicit labels was originally invented to locate or detect particular objects in images in the domain of computer vision \cite{DBLP:conf/aaai/StewartE17}. Researches in multiple disciplines combined the idea (i) to formulate loss functions to include a physics-constrained term \cite{PhysRevLett.126.098302,doi:10.1137/1.9781611975673.63,10.1145/3447548.3467449,doi:10.1177/20414196211073501,vonHahn_Mechefske_2022,Izzatullah2022}, (ii) to eliminate the canonical loss term and totally rely on physical relations \cite{BURWINKEL2022102314,9663012}, or (iii) to solve partial differential equations \cite{ZHU201956,RAISSI2019686,SUN2020112732}.

In this work, we focus on the aspect of timing analysis and propose a novel method building on the intrinsic modularization of detectors. We focus on silicon photomultipliers (SiPM) to generate electronic signals from detectable photons. The motivation of the work is to take full advantage of the universal approximation abilities of NNs while offering an available way to avoid using labelled data and to ease optimization of the model. Compared to some previous works which also used detector signals to estimate time-of-flight (such as \cite{Berg_2018,10038575}), the major innovations and contributions of the paper include:

\begin{itemize}
	\item A practical methodology to use NNs for pulse timing within the conventional nuclear detector dataflow \emph{without} explicit needs of labelled data: a loss function from physical constraints in combination with a \emph{regularizer} automatically guides the NN model to find the optimal solution.
	\item An algorithmic framework to generate the desired \emph{single-channel} time estimates based on an arbitrary time origin by \emph{post-training calibration}.
	\item Experiments with SiPM-based modularized detectors, validation on two experimental datasets, and demonstration of its feasibility and accuracy when applied to detector signals.
\end{itemize}

\section{Methodology}

\subsection{System architecture}

\begin{figure}[htb]
	\centering
	\includegraphics[width=0.98\textwidth]{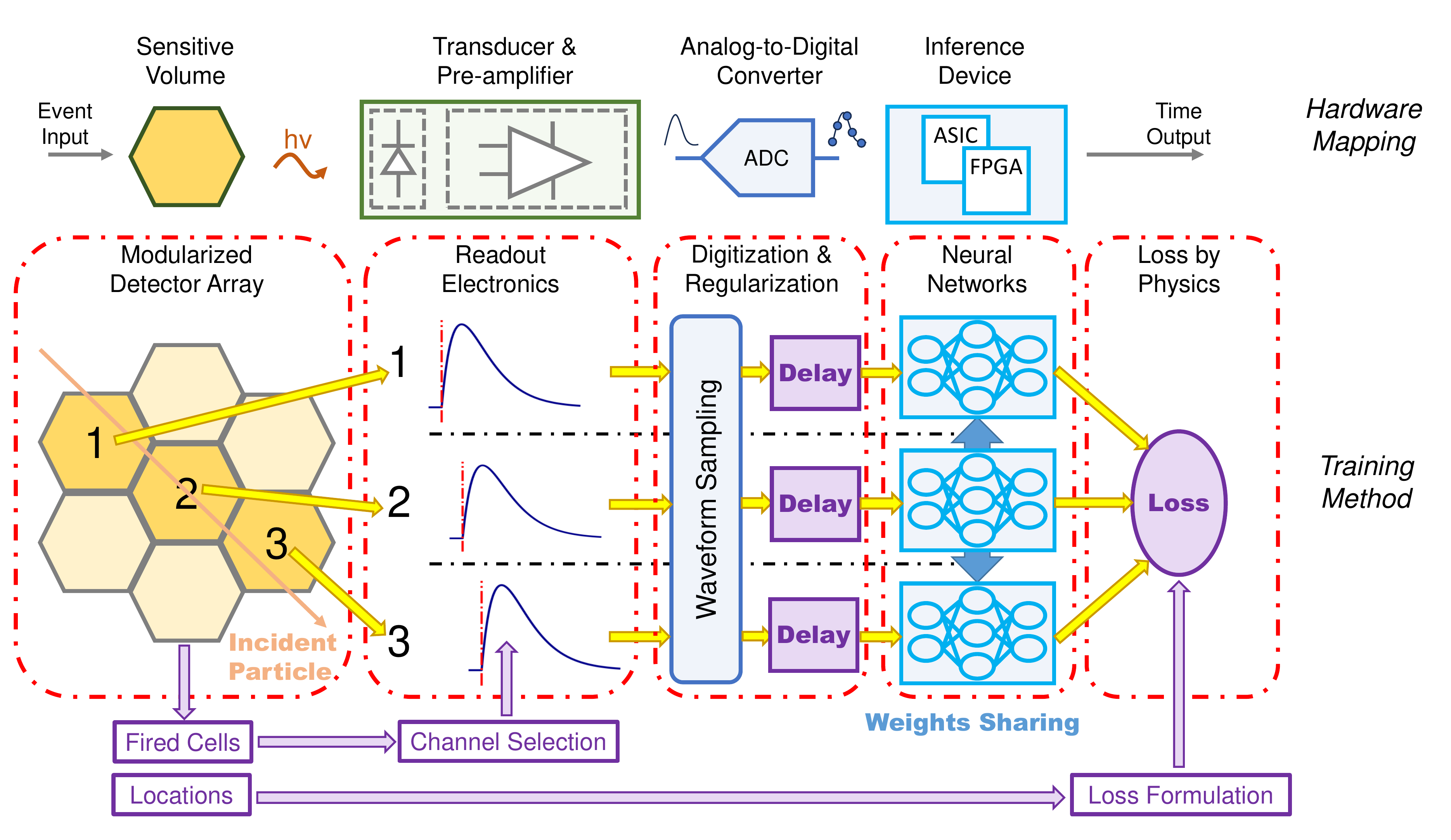}
	\caption{\label{fig:system-diagram} The system diagram of the proposed method based on the conventional nuclear detector dataflow. The modularized detector array provides information of fired cells and their locations for readout electronics to select valid channels and for the training back end to construct physics-constrained formulation of the loss function.}
\end{figure}

The diagram of the proposed system architecture is shown in figure~\ref{fig:system-diagram}. It is based on a paradigm of modern nuclear detectors. The modularized detector array is composed of many detection cells with similar structures and readout channels. In certain conditions, an incident particle will penetrate several detection cells and generate response signals in multiple channels. By monitoring the threshold-crossing amplitudes of signals, it is possible to select fired (being hit with a considerable amplitude) cells and record their locations. The readout electronics keep track of electrical signals from fired cells and send them to following procedures.

The next step is to digitize the analog signals from the readout electronics and to produce series of waveform samples at discrete, equispaced timestamps. As a conventional way of treatment, the digitized waveform samples are ready for feature (time, energy, etc.) extraction by digital logic. To ensure the proper optimization of NNs, we add delay blocks to randomly adjust the time origin of each series of waveform samples and record the adjustment for later use in the loss function (see section~\ref{sec:math-per}).

NNs with shared weights are applied to the outputs of the delay blocks, and each of them is designed to generate a time-of-arrival. Weights sharing is the key to make NNs work in a consistent manner so that a single NN can be used for time prediction with an absolute measure. Here, we intend to incorporate NNs into the system either through offline data analysis or through online processing \cite{AI2020164420,10005128}. Application specific integrated circuits (ASIC) or field programmable gate arrays (FPGA) are candidates for inference devices. Finally, a loss function based on physical constraints, locations of fired cells and delay adjustments is formulated upon the outputs of NNs and back-propagates the residuals to each shared weight with gradient descent optimization.

\subsection{Mathematical perspectives}
\label{sec:math-per}

We start by considering the physical constraints inherited in modularized detectors. The most common kind of constraint is the \emph{linear constraint}. It can be explained as follows: if we denote the moment when the incident particle hits the first cell as $t_0$, the moments when it hits subsequent cells can be denoted as $(t_0 + a_1 t_c,\ t_0 + a_2 t_c,...,\ t_0 + a_{N-1} t_c)$, where $N$ is the number of fired cells, $t_0$ and $t_c$ are unknown variables, and $(a_1,\ a_2,...,\ a_{N-1})$ are known constants. For common geometric structures, linear constraints can be derived very naturally. For example, scintillator bars compactly arranged in an array can be characterized by a linearly constrained timing correlation. In another example, readout signals of different channels in a time projection chamber can also be linearly correlated if the incident particle is not significantly affected by the electric field.

By equating linearly-correlated moments with model predictions, we get:

\begin{equation} \label{equ:linear-cons-sys}
	\bm{A}\left[t_0 \quad t_c\right]^T=\bm{Y}
\end{equation}

\begin{equation*}
	\mathrm{where:} \quad \bm{A} = \left[ {\begin{array}{cc}
		1 & a_0 \\
		1 & a_1 \\
		1 & a_2 \\
		\vdots & \vdots \\
		1 & a_{N-1} \\
	\end{array}} \right] \quad \mathrm{and} \quad \bm{Y} = \left[ {\begin{array}{c}
		y_0 \\
		y_1 \\
		y_2 \\
		\vdots \\
		y_{N-1} \\
	\end{array}} \right]
\end{equation*}

When $N \geq 3$, $a_0$ is fixed at 0, and $a_1, a_2, ..., a_{N-1}$ are not all zeros, the system of linear equations in equation (\ref{equ:linear-cons-sys}) is overdetermined and generally has no accurate solutions. Minimizing the sum of squared residuals $||\bm{Y} - \bm{A}\left[t_0 \quad t_c\right]^T||_2^2$ will yield $\left[\tilde{t}_0 \quad \tilde{t}_c\right]^T = (\bm{A}^T \bm{A})^{-1} \bm{A}^T \bm{Y}$, and the solution can be plugged into the residuals:

\begin{equation} \label{equ:phy-cons-loss}
	I(\bm{Y}) = \left|\left| \left( \bm{I} - \bm{A} (\bm{A}^T \bm{A})^{-1} \bm{A}^T \right) \bm{Y} \right|\right|_2^2
\end{equation}

For convenience, we define $\bm{M} \equiv \bm{I} - \bm{A} (\bm{A}^T \bm{A})^{-1} \bm{A}^T$. It can be easily verified that $\bm{M}$ is a symmetric matrix and $\bm{M}^2 = \bm{M}$. Therefore, equation (\ref{equ:phy-cons-loss}) can be rewritten as:

\begin{equation} \label{equ:phy-cons-loss-trans}
	I(\bm{Y}) = \bm{Y}^T \bm{M}^T \bm{M} \bm{Y} = \bm{Y}^T \bm{M} \bm{Y}
\end{equation}

Next, we will consider how to represent each model prediction $y_i$, the output of individual NNs. $y_i$ depends on waveform samples, which originates from pulse signals generated by the nuclear detector and readout electronics. The time origins and shapes of pulse signals are determined by intrinsic responses of the detector and also extrinsic events. Some assumptions can be made to simplify the analysis:

\begin{enumerate}
	\item The intrinsic responses of the detector are \emph{time-invariant}. In other words, the distributions of pulse signals will not change because of time origins.
	\item The intrinsic responses of different cells have good \emph{uniformity} so that we can use the same pulse shaping function with similar parameters.
	\item The uncertainties of time origins are \emph{homogeneous} so that we can use a single normal distribution to characterize uncertainties of different events.
\end{enumerate}

For assumption (i), it disregards the long-term drift of nuclear detectors due to temperature, ageing, etc., which is a justifiable assumption after each calibration. Assumptions (ii) and (iii) are actually related to the weights sharing and loss formulation of NNs. For theoretical analysis below, they can be weakened, but will also introduce more complex notations and proofs. As a reasonable simplification, we do not consider those variations at the current stage.

Under these assumptions, we write $y_i$ as:
\begin{IEEEeqnarray}{rCl}
	y_i & = & f_{\mathrm{NN}} \left( s_0( 0 t_s - (t_i + \Delta t_i) | \theta_i, n_{i, 0}), ..., s_{K-1}( (K-1) t_s - (t_i + \Delta t_i) |\theta_i, n_{i, K-1}) \right) - \Delta t_i \nonumber \\
	    & \approx & f(t_i + \Delta t_i + \Delta T_i) - \Delta t_i = f(t_0 + a_i t_c + \Delta t_i + \Delta T_i) - \Delta t_i \label{equ:pred-form}
\end{IEEEeqnarray}

\noindent where $f_{NN}(\cdot)$ is the mapping function of NNs, $s_k(\cdot|\theta, n)$ is the k-th waveform sampling point with parameters $\theta$ and noise $n$, $t_s$ is the sampling period, and $\Delta t_i$ is an additional term for regularization. The second line of equation (\ref{equ:pred-form}) is a simplified form of the mapping function under the above assumptions, where $\Delta T_i$ is a random variable representing the \emph{irreducible} spread of time measurements coming from variations of pulse parameters ($\theta$), noise ($n$) and sampling process ($s$).

Finally, we will give a proposition to demonstrate the existence of the optimal solution in common settings:

\begin{proposition}[Sufficient condition for a minimizer] \label{pro:main}
	Assume \( t_0, t_c \) are random variables. \( \Delta t_0, ..., \Delta t_{N-1} \sim \mathcal{N}(0, \sigma_1^2) \) and \( \Delta T_0, ..., \Delta T_{N-1} \sim \mathcal{N}(0, \sigma_2^2) \), both of which are i.i.d random variables, and \( N \geq 3 \). \( I(\bm{Y}) \) is from equation (\ref{equ:phy-cons-loss-trans}) and \( y_i \) in \( \bm{Y} \) is from equation (\ref{equ:pred-form}), where \( a_0 \) is fixed at 0, and \( a_1, a_2, ..., a_{N-1} \) are not all zeros. For \( f : \mathbb{R} \mapsto \mathbb{R} \), if: \[ f(x) = kx + b \quad \mathrm{where:}\quad k = \frac{\sigma_1^2}{\sigma_1^2 + \sigma_2^2},\ \ b = const, \] the following functional: \[ L(f) = \int_{\mathbb{R}^{2N} \times \Omega(t_0, t_c)} I(\bm{Y}) p(\bm{\Delta t}) p(\bm{\Delta T}) p(t_0) p(t_c) \mathrm{d}\bm{\Delta t} \mathrm{d}\bm{\Delta T} \mathrm{d}t_0 \mathrm{d}t_c \] is minimized and takes the minimum value \( \sigma_1^2 \sigma_2^2 \mathrm{\textbf{tr}}(\bm{M}) / ( \sigma_1^2 + \sigma_2^2 ) \).
\end{proposition}

The proof is in \ref{sec:proof}. With this proposition, we guarantee to find a mapping function which is \emph{linearly correlated} with the underlying physical constraint. Some remarks are made as follows:

\begin{enumerate}
	\item The regularization terms ($\Delta t_i$ for $i = 0, 1, ..., N-1$) are essential for the mapping function. It can be seen that, if $\sigma_1 \rightarrow 0$, then $k \rightarrow 0$. This means that the we always find a constant function, which is the "safest" solution under intrinsic uncertainties. To prevent our model from converging to this trivial solution, we must add the regularizers in the formulation of the loss function.
	\item The linear coefficient $k$ is not identity unless there is no intrinsic uncertainties ($\sigma_2 \rightarrow 0$). In real-world cases when intrinsic uncertainties are present, we must perform post-training calibration to achieve identical measures.
	\item The offset $b$ can be an arbitrary constant once the model is well-trained. To make it consistent with time origins in experiments, post-training calibration is needed to make the model free from bias.
\end{enumerate}

Finally, it should be noted that the proposition is only a sufficient condition, which may not always be necessary. However, according to the \emph{Occam's Razor}, a general principle in machine learning, a ``simple'' model (like the linear function) of the underlying physical process is usually to be preferred if the model capacity is appropriate to avoid over-fitting. We will supplement the conclusion with more observations in the section of experimental results.

\subsection{Algorithmic procedure}

\begin{algorithm}
	\caption{Label-free model training and calibration}\label{alg:label-free}
	\begin{algorithmic}
		\Require $\bm{w}_0$: initial weights; $f_{NN}(\cdot;\bm{w})$: NN model; $\eta$: learning rate; $T$: steps for training; $P(\Delta t)$: probability distribution for regularizers; $D$: size of calibration dataset; $G$: steps for linear shift.
		\State \textbf{Model training:}
		\State $\bm{w} \gets \bm{w}_0$ \Comment{Initialize weights for the NN model}
		\For{$i \gets 1, 2, ..., T$} \Comment{Main loop for training}
			\State $N, a_0, a_1, ..., a_{N-1} \gets$ \texttt{ACQUIRE\_GEOMETRY()}
			\State $\bm{s}_0, \bm{s}_1, ..., \bm{s}_{N-1} \gets$ \texttt{ACQUIRE\_SIGNAL()}
			\State Sample $\Delta t_0, \Delta t_1, ..., \Delta t_{N-1} \sim P(\Delta t)$
			\For{$j \gets 0, 1, ..., N-1$}
				\State $\bm{\hat{s}}_j \gets $ \texttt{SHIFT\_WAVEFORM($\bm{s}_j, \Delta t_j$)} \Comment{Shift as regularizers}
				\State $y_j \gets f_{NN}(\bm{\hat{s}}_j;\bm{w}_{i-1}) - \Delta t_j$ \Comment{Apply the NN model}
			\EndFor
			\State $\bm{w}_i \gets \bm{w}_{i-1} - \eta \nabla_{\bm{w}} \bm{I}(y_0, y_1, ..., y_{N-1}; a_0, a_1, ..., a_{N-1})$ \Comment{From equation (\ref{equ:phy-cons-loss-trans})}
		\EndFor
		\State \Return $\bm{w}^* \gets \bm{w}_{T}$
		\State
		\State \textbf{Model calibration:}
		\State $i \gets 0$, $S_{\mathrm{eval}} \gets \O$
		\While{$i < D$} \Comment{Gather calibration dataset}
			\State $\bm{s}_0, \bm{s}_1, ..., \bm{s}_{N-1} \gets$ \texttt{ACQUIRE\_SIGNAL()}
			\State $S_{\mathrm{eval}} \gets S_{\mathrm{eval}} \bigcup \{\bm{s}_0, \bm{s}_1, ..., \bm{s}_{N-1}\}$
			\State $i \gets i + N$
		\EndWhile
		\For{$i \gets 1, 2, ..., D$}
			\State $\bm{s}_i \gets S_{\mathrm{eval}}[i]$
			\For{$j \gets 0, 1, ..., G-1$}
				\State $\bm{\hat{s}}_{i, j} \gets $ \texttt{SHIFT\_WAVEFORM($\bm{s}_i, j$)}
				\State $z_{i,j} \gets f_{NN}(\bm{\hat{s}}_{i, j};\bm{w}^*)$
			\EndFor
			\State $k_i, b_i \gets $ \texttt{LINEAR\_FITTING($z_{i, 0}, z_{i, 1}, ..., z_{i, G-1}$)} \Comment{Compute slope and intercept}
		\EndFor
		\State $k^*, b^* \gets$ \texttt{COMPUTE\_MEAN($k_1, k_2, ..., k_D; b_1, b_2, ..., b_D$)}
		\State \Return $f_{NN}^*(\cdot; \bm{w}^*) \gets (f_{NN}(\cdot; \bm{w}^*) - b^*)/k^*$ \Comment{Normalize the model}
	\end{algorithmic}
\end{algorithm}

Based on the above theory and architecture, we propose the label-free model training and calibration procedure, as shown in Algorithm~\ref{alg:label-free}.

For training, weights of NN are initialized at first. Then in the main loop, the weights are optimized with the loss function constrained by physical information. We use \texttt{ACQUIRE\_GEOMETRY()} to represent the process of recording the number of fired cells and their locations, and \texttt{ACQUIRE\_SIGNAL()} to represent the process of reading out waveform \emph{samples} from each fired cell to form \emph{examples}\footnote{Throughout the paper, we use the term ``sample'' to refer to a scalar value, and the term ``example'' to refer to a vector of the ADC-sampled waveform composed of samples, or a time-related group of vectors, depending on the context.}. Regularization by random shift is applied to the examples. Instead of sampling the random shift from a continuous Gaussian distribution, we actually draw samples from a discrete Bernoulli distribution with equal probabilities of taking $+1$ and $-1$\footnote{It can be verified that Proposition~\ref{pro:main} still holds for this distribution.}. The shifted examples generated from the \texttt{SHIFT\_WAVEFORM()} process are propagated throughout NN, and the amount of random shift is subtracted from the output. Finally, a stochastic gradient descent is applied to the physics-constrained loss function to update the weights of NN. This procedure is repeated until we reach the maximum steps for training.

For calibration, the first loop is to gather enough examples to form a calibration dataset regardless of their geometric information. In the next nested loop, each example from the calibration dataset is shifted multiple times with the minimum step. Then these shifted examples are propagated throughout NN with well-trained weights to get multiple outputs. The process \texttt{LINEAR\_FITTING()} is applied to the outputs and generates a slope and an intercept for this example in the calibration dataset. Finally, the process \texttt{COMPUTE\_MEAN()} is applied to all the slopes and intercepts to determine their optimal values. The NN model with well-trained weights is shifted and scaled to generate the normalized model.

With the normalized model, it is ready to re-calibrate the output based on the actual sampling period of ADC at the front end, and on a required time origin synchronous to the experimental detector system.

\section{Experimental results} 

\subsection{A toy experiment}

In this section, we consider a \emph{degenerated} case: the predicted time from two cells should be equal. In this case, $N$ equals to 2, and the matrix $\bm{A}^T\bm{A}$ will be singular and no longer invertible. However, we can still find a symmetric matrix $\bm{M}$:

\begin{equation} \label{equ:toy-m}
	\bm{M} = \frac{1}{2} \left[ {\begin{array}{cc}
			1 & -1 \\
			-1 & 1 \\
	\end{array}} \right]
\end{equation}

\noindent which satisfies $\bm{M}^2 = \bm{M}$ and fully agrees with the constraint. In this case, Proposition~\ref{pro:main} still holds with the only difference being the value of $N$.

\subsubsection{Experimental setup}

\begin{figure}[htb]
	\centering
	\begin{subfigure}[b]{0.6\textwidth}
		\centering
		\includegraphics[width=\textwidth]{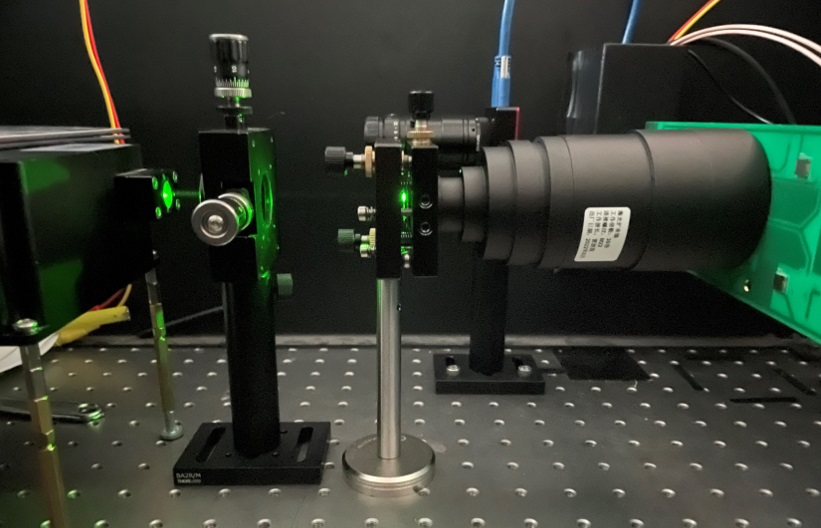}
		\caption{Photograph}
	\end{subfigure}
	\begin{subfigure}[b]{0.75\textwidth}
		\centering
		\includegraphics[width=\textwidth]{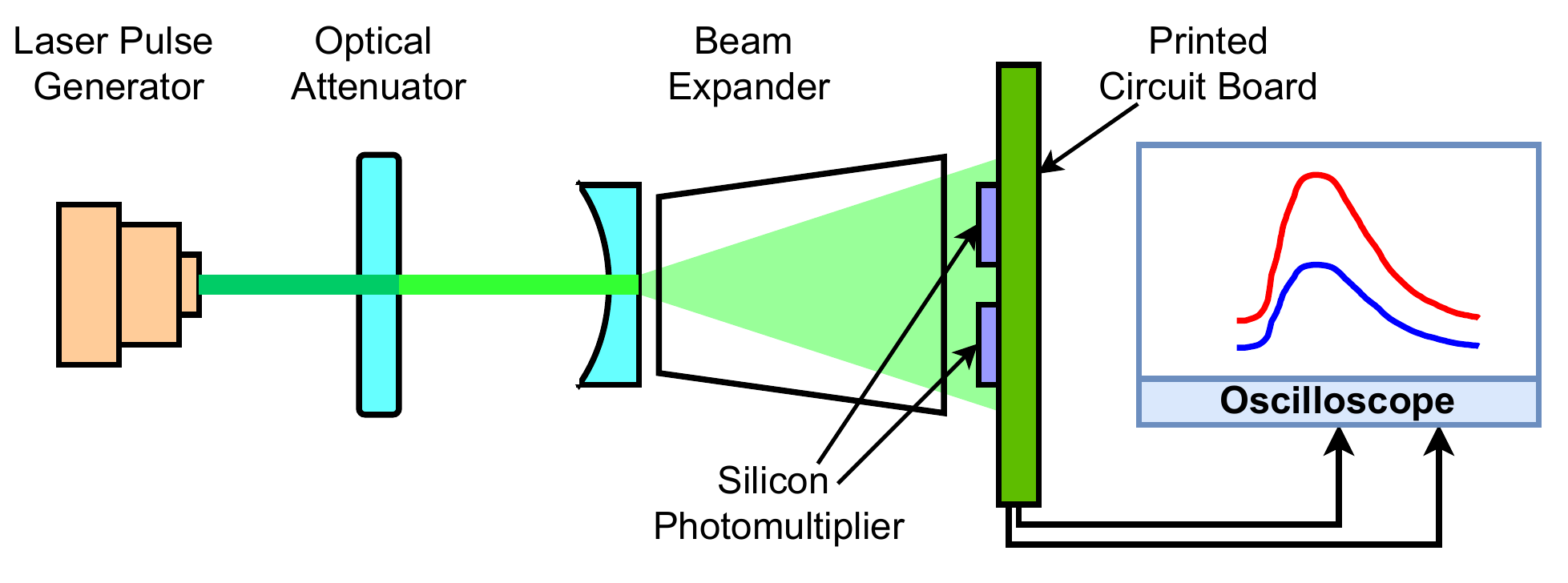}
		\caption{Diagram}
	\end{subfigure}
	\caption{\label{fig:toy-exp-setup} Devices and equipment for the toy experiment.}
\end{figure}

The setup for the toy experiment is shown in figure~\ref{fig:toy-exp-setup}. A rapid laser pulse generator (GZTech YFL-PP-1.5-GR) produces green lights with 532-\si{nm} wavelength periodically at 100 \si{kHz}. An optical attenuator (Thorlabs NE40A) shrinks the intensity of the laser lights (effectively the number of photons) to a reasonable level. After that, a beam expander (YZL D60-M30) spreads the arrived photons uniformly to two SiPMs (Hamamatsu S13360-6025PE) devices soldered onto a printed circuit board. The SiPMs transduce the light signals into electronic pulses, which are conditioned by pre-amplification circuits and read out by a 10-\si{GSPS} digital oscilloscope (LeCroy WaveRunner8254). The experiment is conducted in a black box to avoid ambient light.

Though not perfectly aligned, the arrival time of a few photons at two SiPMs could be considered the same, and the intrinsic time resolution is largely determined by the electronics. Our goal is to produce a model capable of predicting the arrival time from waveform samples digitized by the oscilloscope, and also to find out the intrinsic time resolution of the analog channels.

We implement the NN model in Keras \cite{chollet2015} with the TensorFlow \cite{DBLP:journals/corr/AbadiABBCCCDDDG16} back end. The \emph{base model} ($f_{\mathrm{NN}}(\cdot)$ in equation (\ref{equ:pred-form})) is a classic convolutional neural network (CNN) comprising several one-dimensional convolution layers and several fully-connected layers. The details of the architecture are described in \ref{sec:detail-nn-arch-toy}. The network capacity is relatively small compared to those used in computer vision. We use CNN as a representative of NN models because it reaches near-optimal performance in the toy experiment, while more variants of NNs are discussed in section~\ref{sec:nica-mpd}.

The input to the base model is a series of 2048 waveform samples with an interval of 0.1 \si{ns}, from either the first channel or the second channel. To ensure the proper functioning of the physical loss function, an appropriate amount of randomness is vital for NN models to converge to meaningful mapping and generalize to unseen examples. The original data have already exhibited intrinsic randomness to some extent (see figure \ref{fig:data-inspect}). To enhance the variations, an additional shift of the sampling window is applied to each example (see \ref{sec:detail-conf} for more details).

The output of the calibrated base model is a continuous value of the predicted arrival time with regard to the time origin of the sampling session\footnote{In a large detector system, the time prediction will first be aligned to the ADC clock domain, and then synchronized to a sophisticated clock distribution system.}. Two base models with weight sharing are built for two analog channels. For regularization, a random shift of $+1$ or $-1$ interval applies separately to each channel input and works as $\Delta t_i$ in equation (\ref{equ:pred-form}). We use the physical loss function in equation (\ref{equ:phy-cons-loss-trans}) to train the NN model with the Adam \cite{DBLP:journals/corr/KingmaB14} optimization algorithm. More details about the configurations are described in \ref{sec:detail-conf-toy}.

\subsubsection{Main results}

\begin{figure}[htb]
	\centering
	\begin{subfigure}[b]{0.45\textwidth}
		\centering
		\includegraphics[width=\textwidth]{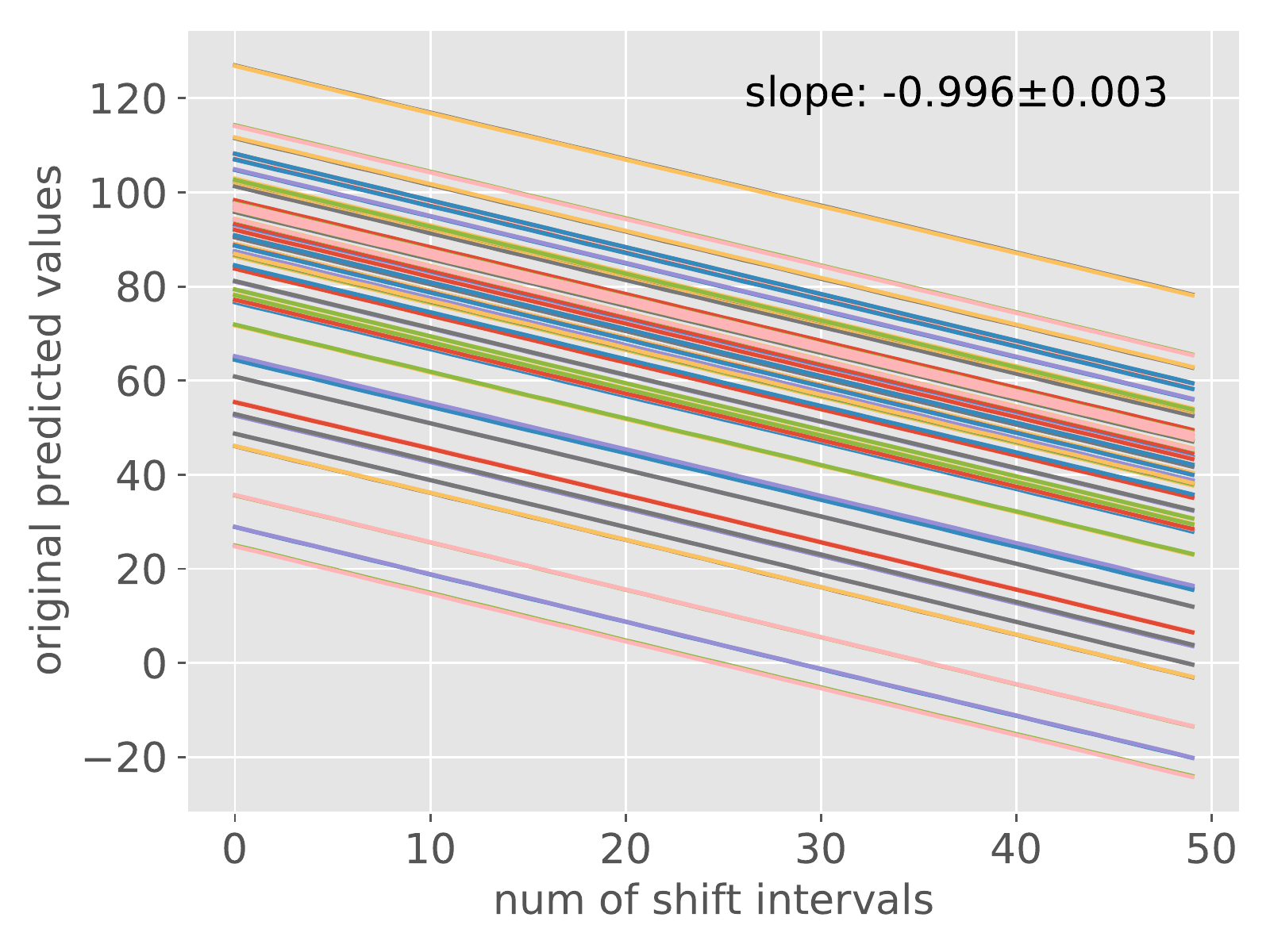}
		\caption{Test on first 100 examples.}
		\label{fig:s-toy-avg-slope}
	\end{subfigure}
	\begin{subfigure}[b]{0.45\textwidth}
	\centering
	\includegraphics[width=\textwidth]{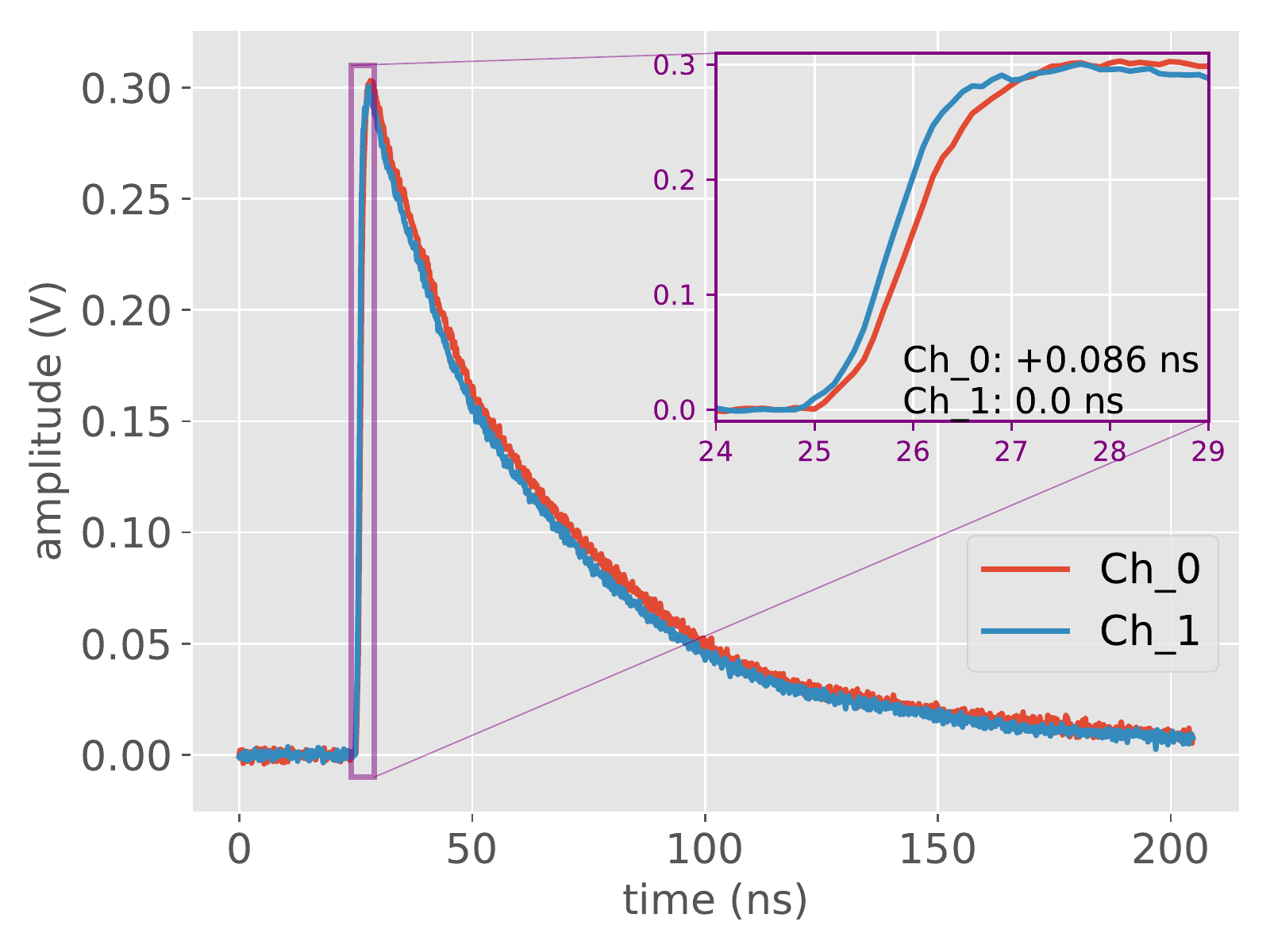}
	\caption{An example of prediction.}
	\label{fig:s-toy-pred-example}
	\end{subfigure}
	\begin{subfigure}[b]{0.45\textwidth}
		\centering
		\includegraphics[width=\textwidth]{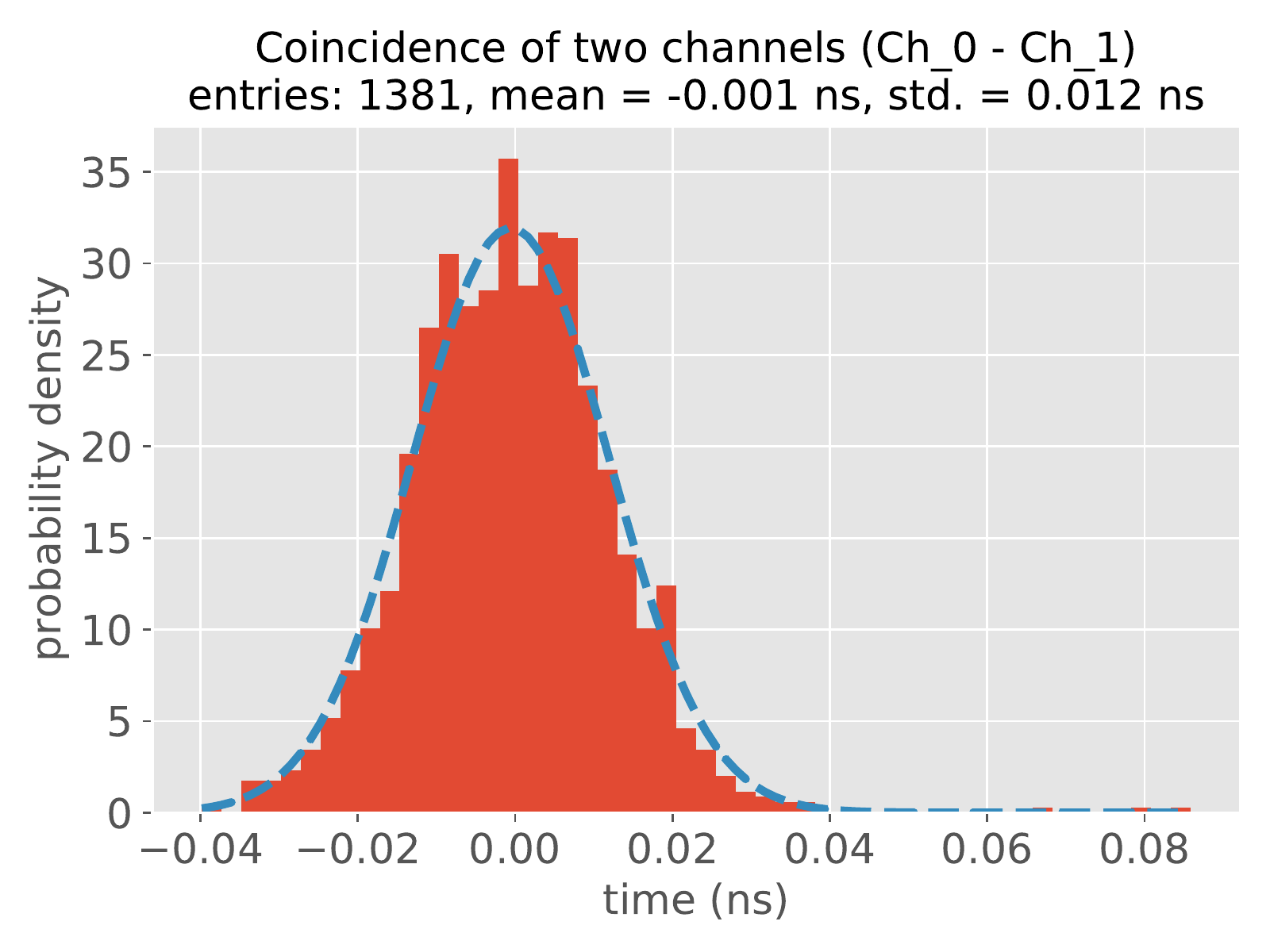}
		\caption{Histogram of timing differences.}
		\label{fig:s-toy-hist-diff}
	\end{subfigure}
	\caption{Demonstration of NN predictions in the toy experiment. (a) To calibrate the model, all examples are equidistantly shifted in a given range and fed to the base model for prediction. The average slope is computed by all slopes from linear fitting. Here, we plot the original (uncalibrated) model outputs of first 100 examples, which show very high linearity and consistency. The calibration will involve normalizing the model to the scale of sampling intervals and rescaling with the actual ADC interval (0.1 \si{ns}). (b) A segment showing the rising edge of an example in the test dataset, on its original scale, and associated timing prediction results. (c) Histogram of all timing differences in the test dataset when no low-pass filtering is performed.}
	\label{fig:s-toy-demo}
\end{figure}

We follow Algorithm~\ref{alg:label-free} to train the NN model for optimal weights, and to calibrate the model so that the scale of network outputs is adjusted to match the actual ADC interval. Figure~\ref{fig:s-toy-avg-slope} shows the correlation between the original (uncalibrated) outputs of NN and the number of shift intervals. While the slopes are negative as a result of the shift method, it is straightforward to transform them into positive numbers for Algorithm \ref{alg:label-free} to be applied. By examining this figure, we find that the linear dependency of predicted values on shift intervals is almost exact, and the plotted 100 examples are very consistent when measuring their slopes. Besides, the linear range of predicted values is reasonably wide to cover all examples in the dataset.

In figure~\ref{fig:s-toy-pred-example}, we select the rising edge of an example in the test dataset on its original scale with physical meanings. Since timing differences in most examples are very tiny, for visual convenience, we show an extreme case with the maximum timing difference (the rightmost point in figure~\ref{fig:s-toy-hist-diff}). It can be seen that the waveform is clean with a low noise level. The predicted values give a correct timing relation ($+0.086$ \si{ns}) corresponding to this example.

The overall performance on the test dataset is presented as a histogram of calibrated timing differences between two channels, shown in figure~\ref{fig:s-toy-hist-diff}. Apart from several outliers, nearly all values locate near the origin and conform to a Gaussian distribution. It should be noted that the standard deviation of the Gaussian distribution is 12.5 \si{ps} (equivalent to 8.8 \si{ps} for a single channel), which is a considerably low resolution level, showing the good performance of the experimental system and also the NN model.

\begin{figure}[htb]
	\centering
	\includegraphics[width=0.75\textwidth]{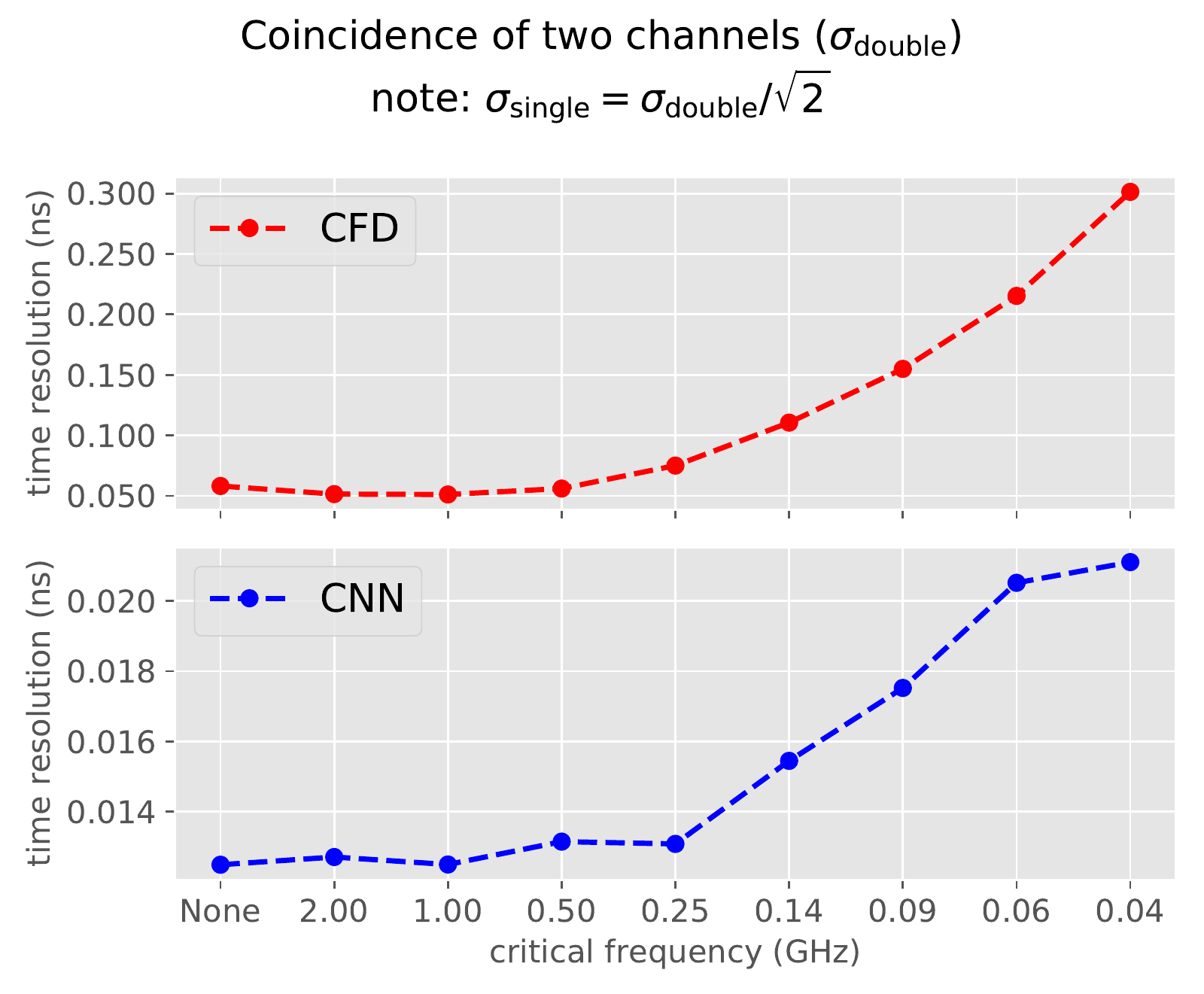}
	\caption{Comparison of the digital constant fraction discrimination (CFD) and the convolutional neural network (CNN). The time resolution is the standard deviation of timing differences, plotted versus variable critical frequencies of low-pass filters.}
	\label{fig:s-toy-cfd-cnn}
\end{figure}

A comparison is conducted between the NN model and the digital constant fraction discrimination (CFD) \cite{FALLULABRUYERE2007247} (a traditional timing method in nuclear electronics, see \ref{sec:detail-conf-toy} for details). To simulate the analog channel, we preprocess the waveform with a second-order low-pass filter. Figure~\ref{fig:s-toy-cfd-cnn} shows the comparison under different critical frequencies of the analog channel. It can be seen that CNN performs consistently better than digital CFD. The best performance of CFD lies in $1 \sim 2$ \si{GHz} critical frequency, because noise is more intense when the analog channel has higher bandwidth, which degrades the performance of CFD. However, CNN can still reach good performance even when no low-pass filtering is present, due to the intricate network architecture which works as a non-linear filter and exploits the information in the input waveform samples as much as possible.

It should be noted that the optimal critical frequency will change if we decrease the sampling rate of the digitizer. At a lower sampling rate, in order to gather enough sampling points at the fast-rising edge (which is important for pulse timing), a low-pass filter with narrower passband and smaller critical frequency is preferred. This will rule out more high-frequency noise as well as rapid-changing details of the electronic pulse. When the noise is not a dominant limiting factor, we expect the timing resolution to keep the same order of magnitude \cite{Ai_2021}.

\subsubsection{Robust against concept drift}

\begin{figure}[htb]
	\centering
	\includegraphics[width=0.75\textwidth]{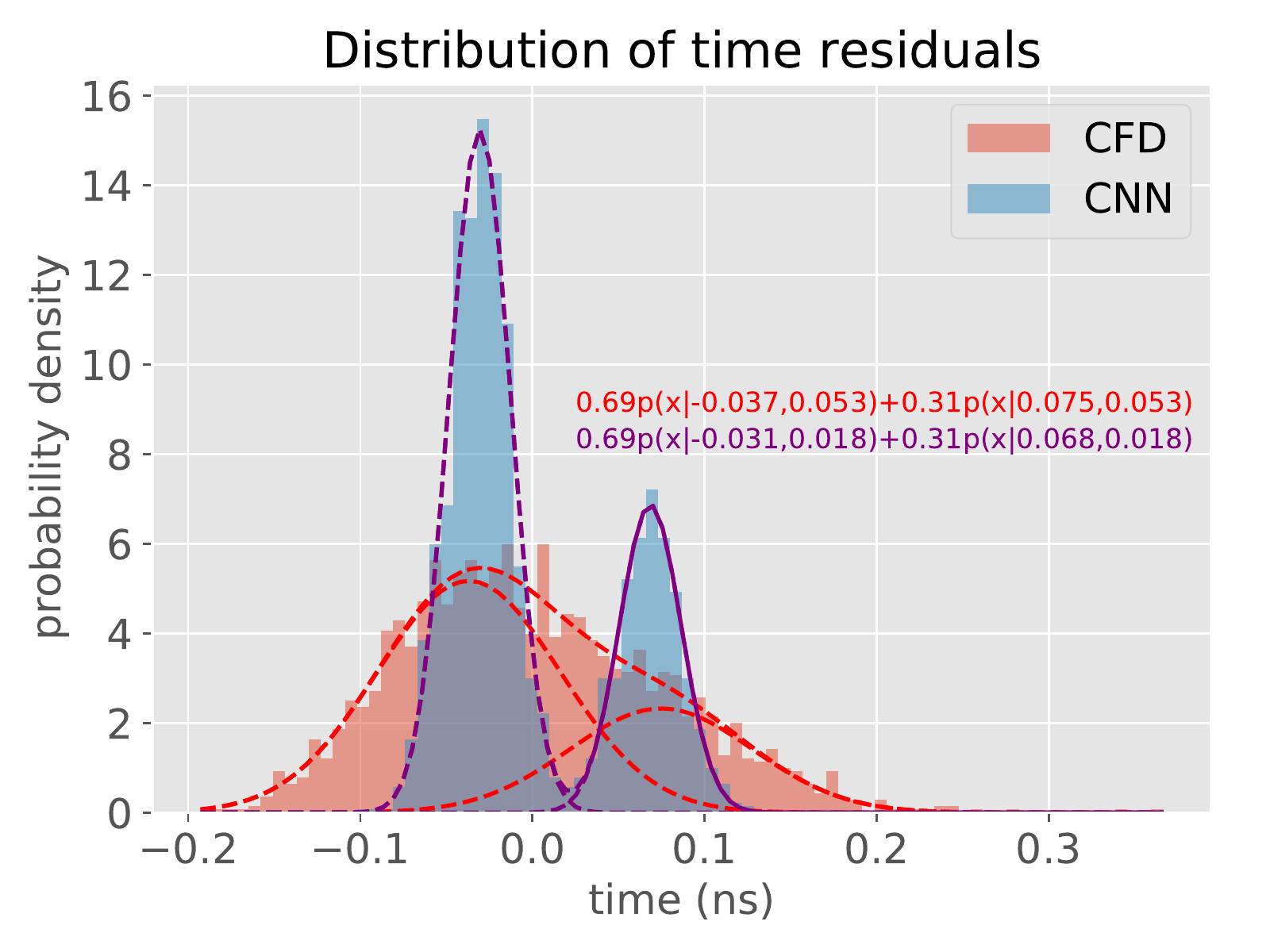}
	\caption{Histograms of timing differences by CFD and CNN when two distinct modalities exist in the dataset. We fit the histograms to a Gaussian mixture with two components. CNN gives enough resolution to distinguish two peaks, while CFD is unable to display such feature.}
	\label{fig:s-toy-two-gauss}
\end{figure}

During long time of data capturing with the oscilloscope, the real-time correspondence of two analog channels may change\footnote{This phenomenon is somewhat discovered by accidence. Nevertheless, we think it is helpful to explain the merits of the proposed method.}. In this condition, the premise of the experiment has changed, and we call it a \emph{concept drift} in machine learning: the arrival time of two analog channels can \emph{not} be regarded as the \emph{same} for all examples in the dataset. By analysing the loss function, the concept drift introduces a bias term when predictions are assumed to be accurate. Though theoretical proof of the property under this condition is non-trivial and lies beyond the scope of the paper, empirical demonstrations can be made with quantitative results.

In figure~\ref{fig:s-toy-two-gauss}, we visualize the distribution of timing residuals when the actual difference of arrival time between two analog channels takes two distinct values. To our surprise, the NN model can be trained to provide accountable predictions against concept drift, and achieve fairly well performance. To be more specific, we fit the histograms to the following function:

\begin{equation}
	f(x; a, \mu_1, \mu_2, \sigma) = a \cdot p(x; \mu_1, \sigma) + (1 - a) \cdot p(x; \mu_2, \sigma)
\end{equation}

\noindent where $p(x; \mu, \sigma)$ is the probability density of a Gaussian distribution. The fitting results are exhibited on the right side of the figure. For CNN, the standard deviation of each Gaussian component is $\sim$18 \si{ps}, which is slightly worse than the single modality condition but still remarkable. By examining the portion parameter ($a$) and mean parameters ($\mu_1, \mu_2$), it can be seen that the means of two components are at a distance of approximately one sample interval (0.1 \si{ns}), and the combined mean value is located near the origin. For CFD, it is unable to separate two Gaussian peaks due to its relatively poor resolution. The standard deviation ($\sim$53 \si{ps}) shows accordance with the single modality condition regardless of concept drift.

For CNN, it is interesting to draw an analogy with solving a second-order minimizing problem with bias in one of two components. Here, the major difference is that the variable originally in the analogical problem is itself a mapping function with its own inputs and trainable parameters, which makes the analytical solution intractable.

\subsection{Electromagnetic calorimeter at NICA-MPD}
\label{sec:nica-mpd}

In this section, we study a detector in a real-world high energy physics experiment -- the electromagnetic calorimeter (ECAL) \cite{NICA-MPD-ECAL-TDR} of the Multi-Purpose Detector (MPD) \cite{NICA-MPD-CDR} at the NICA collider. ECAL is a huge and fully modularized detector assembled in several hierarchies (tower, module, half sector, sector). The basic unit for assembling is the tower, which is a shashlik-type sampling calorimeter composed of alternate lead and scintillator plates stacked one kind after another. In each tower, 16 wavelength shift fibres penetrate the bulk of the detection area and guide the scintillation lights onto a SiPM, which will then be read out by front-end electronics. A combination of $2 \times 8$ towers form a module. The modules from different sectors differ in their shapes. In this experiment, we use a module from sector 8 (the sector in the outermost place of ECAL) as the subject.

When a particle with high energy (such as muons in the cosmic ray) enters the tower, it will generate an electromagnetic cascade, in which the absorber gradually deposits the energy and the scintillator transforms it to detectable lights transmitted through the fibres. Reaching sub-\si{ns} timing resolution is meaningful for ECAL to work as an auxiliary time-of-flight device for better particle identification.

\subsubsection{Experimental setup}

\begin{figure}[htb]
	\centering
	\begin{subfigure}[b]{0.6\textwidth}
		\centering
		\includegraphics[width=\textwidth]{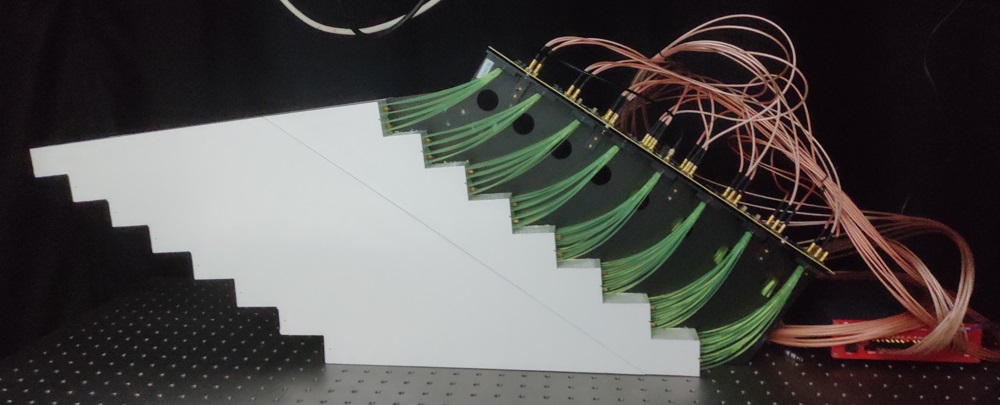}
		\caption{Photograph of the ECAL module}
	\end{subfigure}
	\begin{subfigure}[b]{0.75\textwidth}
		\centering
		\includegraphics[width=\textwidth]{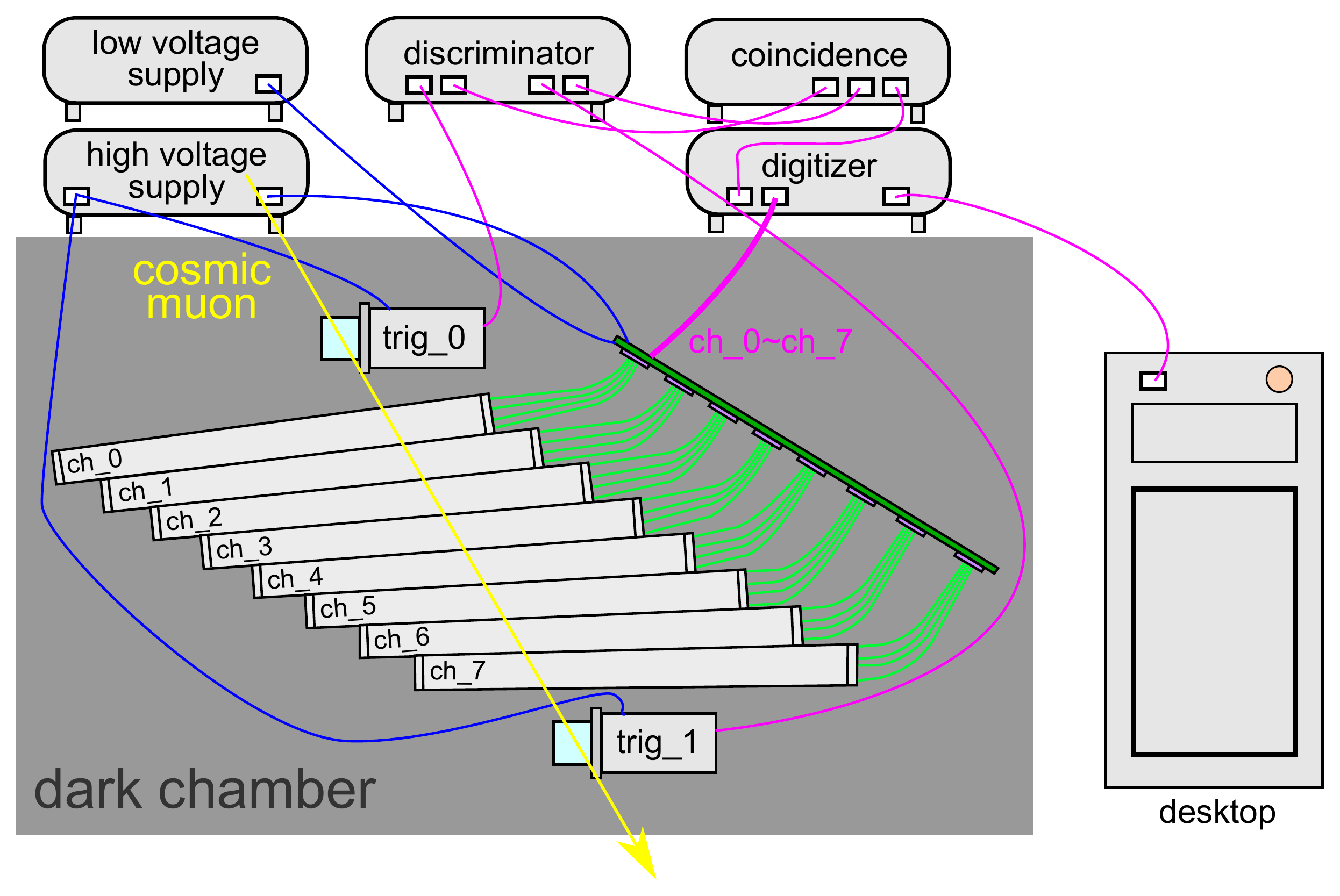}
		\caption{Diagram}
	\end{subfigure}
	\caption{\label{fig:ecal-exp-setup} Devices and equipment for the ECAL experiment.}
\end{figure}

The setup for the ECAL experiment is shown in figure~\ref{fig:ecal-exp-setup}. The ECAL module is put horizontally in a dark chamber. Two trigger devices (scintillator with photomultiplier) are set above and beneath the ECAL module to capture cosmic muons passing through 8 towers of the ECAL module (the other 8 towers are not used). The outputs of two triggers are connected to the leading edge discriminator (CAEN N840) to generate step signals, which are reduced to a single pulse by the coincidence logic unit (CAEN N455). The coincidence pulse is transmitted as the trigger input to the switched capacitor digitizer (CAEN DT5742). The digitizer also takes the 8 analog channels of the ECAL module as inputs and samples the waveform at a rate of 5 \si{GSPS}. The digital data are sent to a desktop computer for storage. High voltage supply is provided to two triggers and the electronics of the ECAL module, and low voltage supply is provided to the electronics.

In most captured events, the cosmic muon leaves energy deposits in every passed tower. If we only consider the median interaction point between the particle and each tower, the time of incidence will exhibit a highly linearly-correlated relation between these towers\footnote{The relation is geometrically accurate except for the slightly different fibre length of individual towers. However, it will not influence the training process if we consider it as a constant bias to the loss function.}, as long as the trajectory of the cosmic muon reaches two triggers. However, the actual process of energy deposition, scintillation and photon transmission is non-trivial for the detector, and there is obvious electronic noise relative to the signal amplitude. These factors make the linear correlation only a target approximation. Nevertheless, by utilizing it to form a loss function, we can train a model to work as an arrival time predictor and find out the intrinsic time resolution of the ECAL module at the same time.

The NN model is implemented with settings similar to the toy experiment. In this section, we construct the base model with three types of NNs: the fully-connected (FC) network, the convolutional neural network (CNN) and the recurrent neural network (RNN) with long short-term memory (LSTM). The details of their architectures are described in \ref{sec:detail-nn-arch-ecal}. The input to the base model is a series of 800 waveform samples with an interval of 0.2 \si{ns}, from one of the eight analog channels. More details about of the configurations are described in \ref{sec:detail-conf-ecal}.

\subsubsection{Main results}

\begin{figure}[htb]
	\centering
	\begin{subfigure}[b]{0.45\textwidth}
		\centering
		\includegraphics[width=\textwidth]{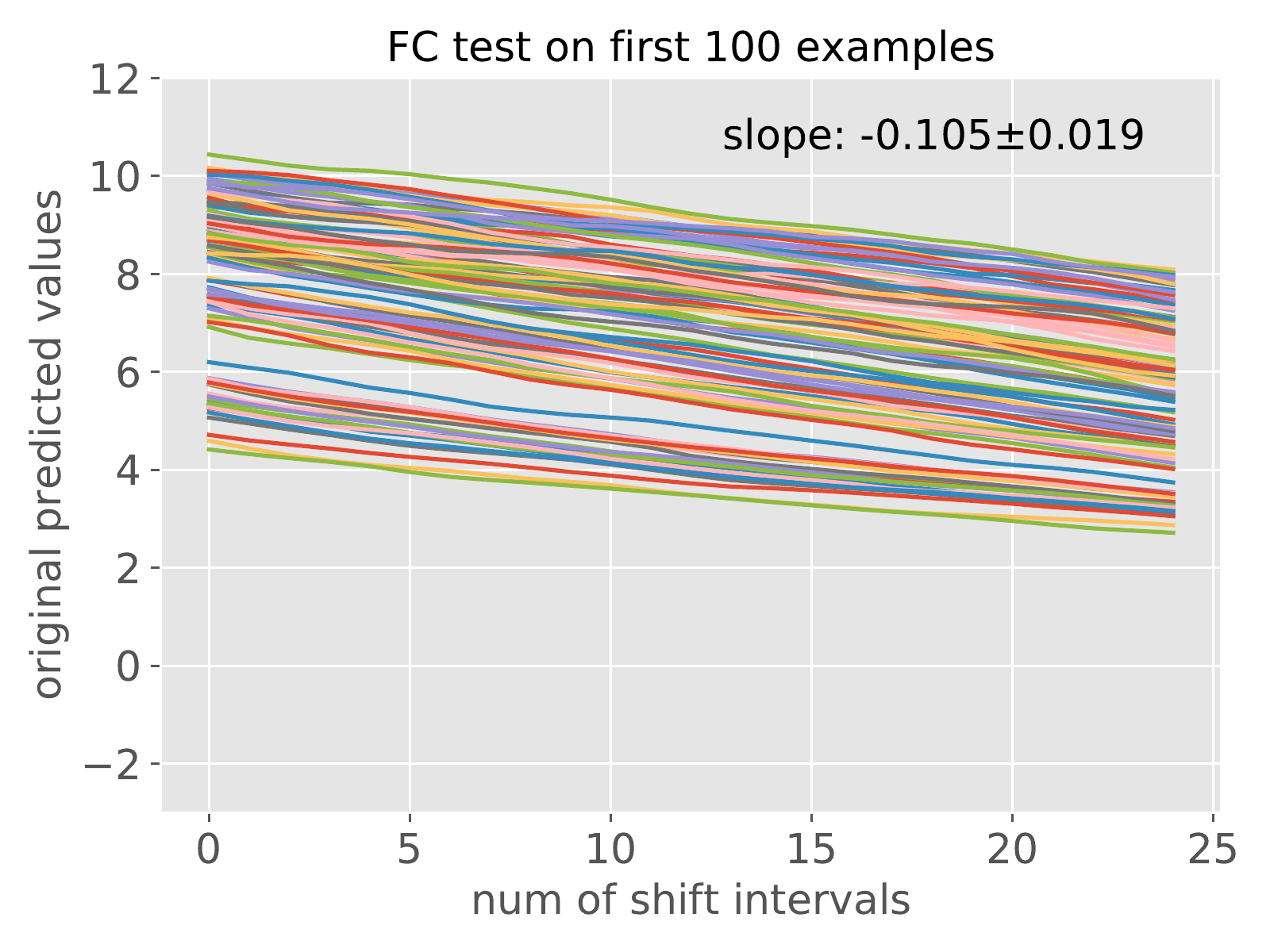}
		\caption{FC}
		\label{fig:fc-avg-slope}
	\end{subfigure}
	\begin{subfigure}[b]{0.45\textwidth}
		\centering
		\includegraphics[width=\textwidth]{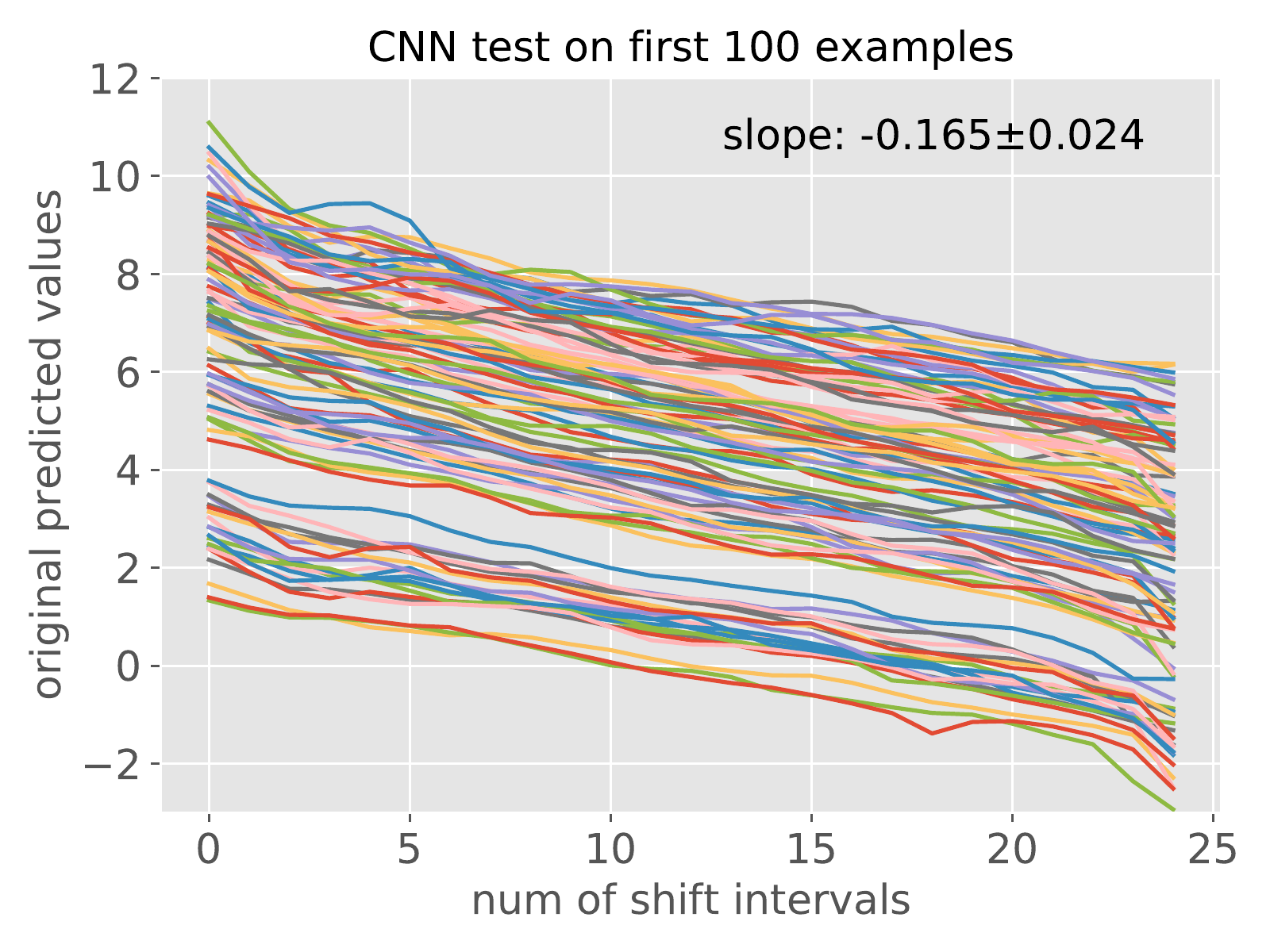}
		\caption{CNN}
		\label{fig:cnn-avg-slope}
	\end{subfigure}
	\begin{subfigure}[b]{0.45\textwidth}
		\centering
		\includegraphics[width=\textwidth]{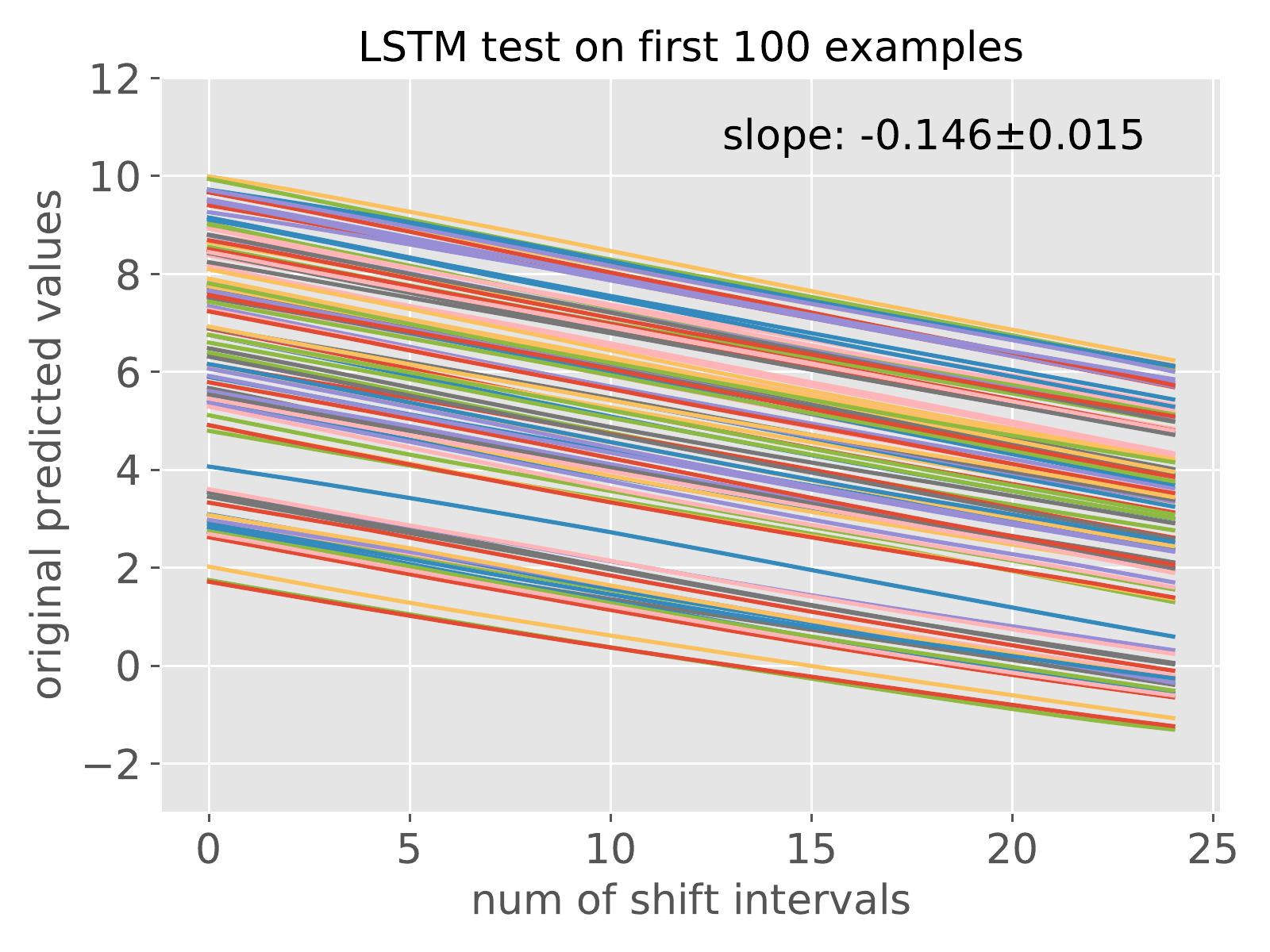}
		\caption{LSTM}
		\label{fig:lstm-avg-slope}
	\end{subfigure}
	\caption{Demonstration of the calibration process (showing only the first 100 examples) with equidistantly shifted examples and their original (uncalibrated) model outputs (See figure \ref{fig:s-toy-avg-slope} for a reference): (a) Fully-connected network; (b) Convolutional neural network; (c) Recurrent neural network with long short-term memory.}
	\label{fig:s-basic-avg-slope}
\end{figure}

\begin{figure}[htb]
	\centering
	\begin{subfigure}[b]{0.45\textwidth}
		\centering
		\includegraphics[width=\textwidth]{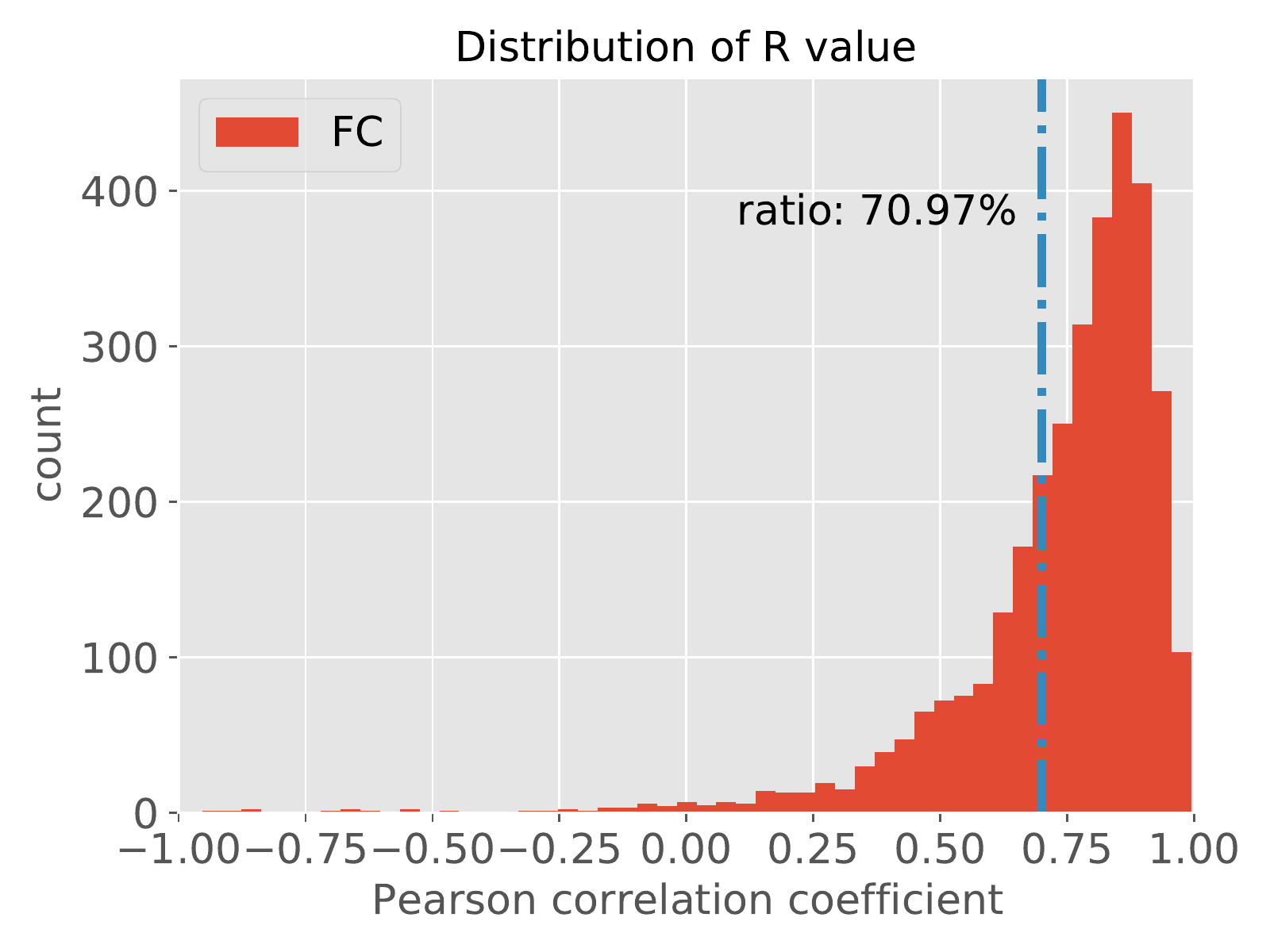}
		\caption{FC}
		\label{fig:fc-r-value}
	\end{subfigure}
	\begin{subfigure}[b]{0.45\textwidth}
		\centering
		\includegraphics[width=\textwidth]{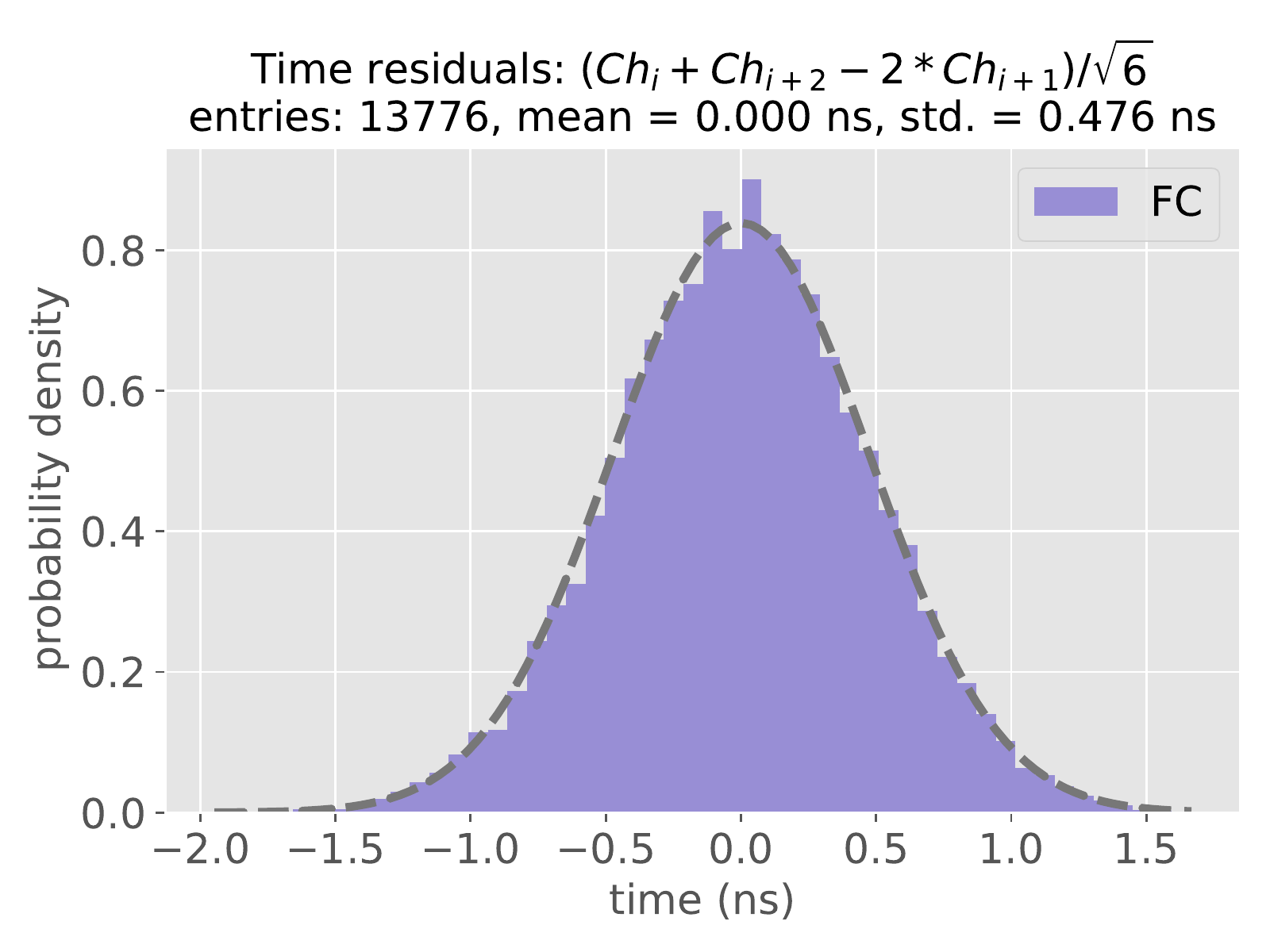}
		\caption{FC}
		\label{fig:fc-hist-on-thresh}
	\end{subfigure}
	\begin{subfigure}[b]{0.45\textwidth}
		\centering
		\includegraphics[width=\textwidth]{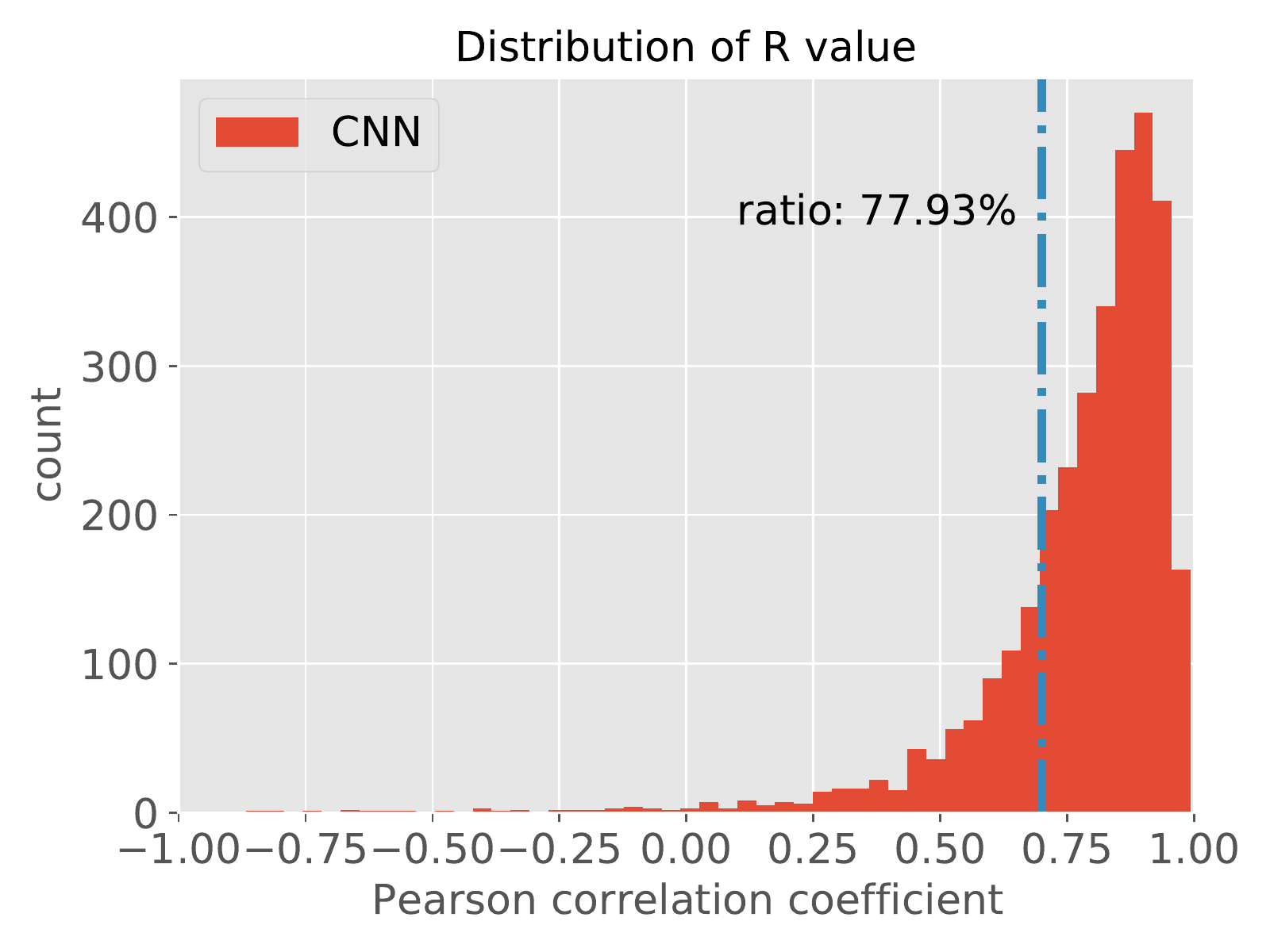}
		\caption{CNN}
		\label{fig:cnn-r-value}
	\end{subfigure}
	\begin{subfigure}[b]{0.45\textwidth}
		\centering
		\includegraphics[width=\textwidth]{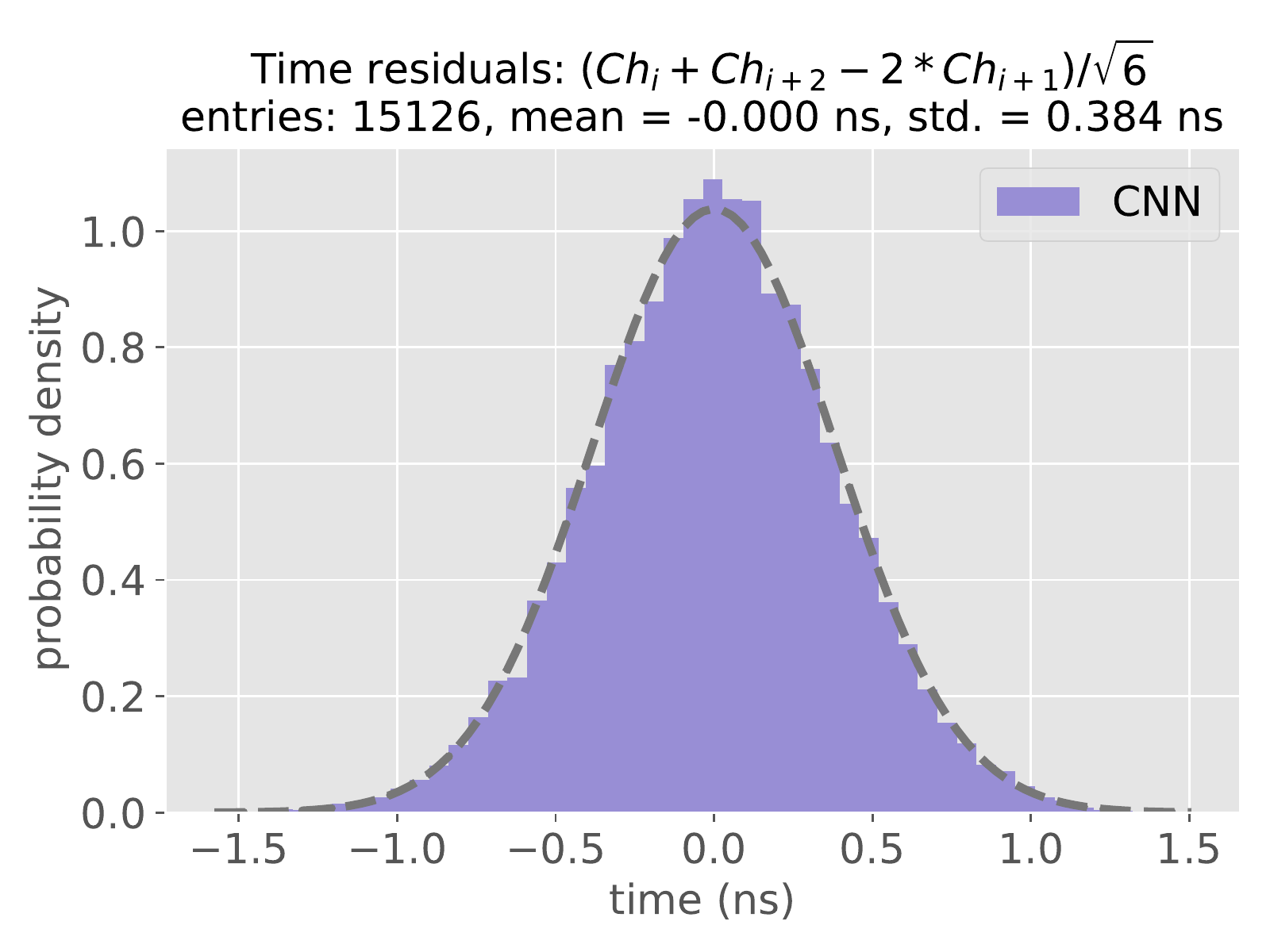}
		\caption{CNN}
		\label{fig:cnn-hist-on-thresh}
	\end{subfigure}
	\begin{subfigure}[b]{0.45\textwidth}
		\centering
		\includegraphics[width=\textwidth]{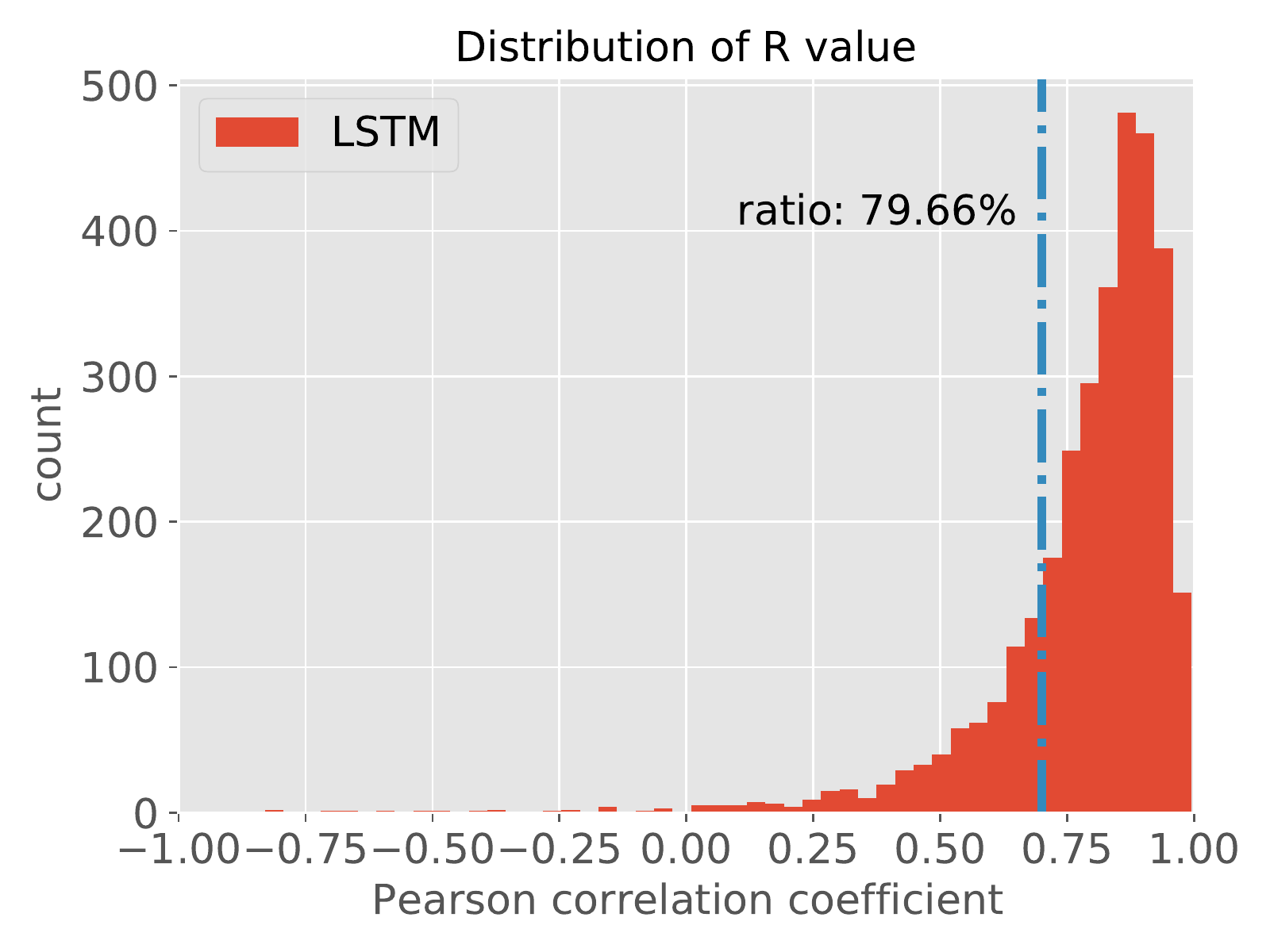}
		\caption{LSTM}
		\label{fig:lstm-r-value}
	\end{subfigure}
	\begin{subfigure}[b]{0.45\textwidth}
		\centering
		\includegraphics[width=\textwidth]{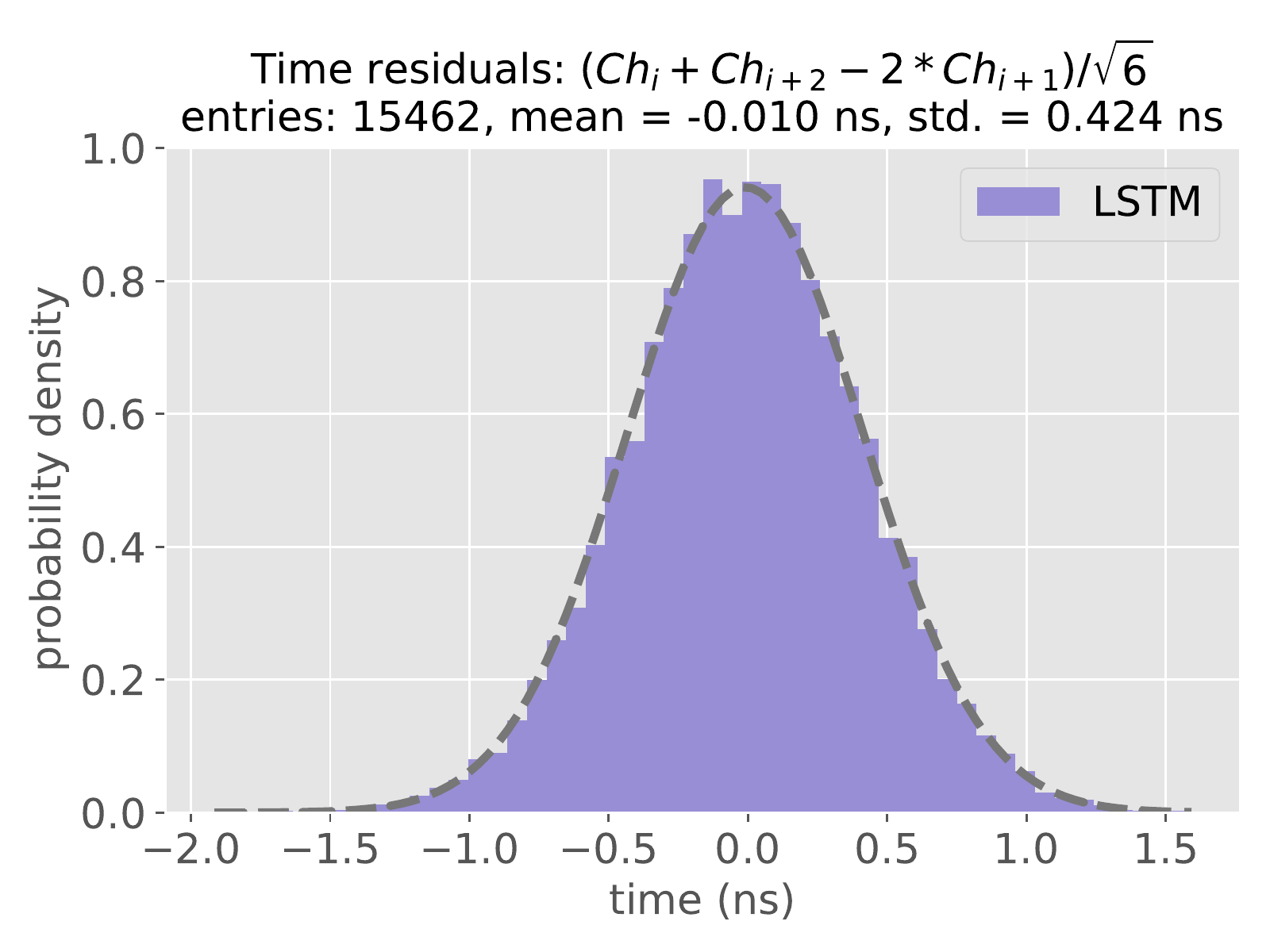}
		\caption{LSTM}
		\label{fig:lstm-hist-on-thresh}
	\end{subfigure}
	\caption{(a)(c)(e) Histograms of the R value (Pearson correlation coefficient) distribution by linearly fitting NN predictions of 8 time-correlated analog channels, in the test dataset. The threshold (0.7) and the ratio of examples above the threshold are indicated. (b)(d)(f) Histograms of time residuals when examples with linearity above the threshold (0.7) are selected. We fill the histograms with triples of adjoined analog channels: $(Ch_{i} + Ch_{i+2} - 2 * Ch_{i+1}) / \sqrt{6}$.}
	\label{fig:nn-hist}
\end{figure}

Similarly to the toy experiment, we follow Algorithm~\ref{alg:label-free} to train and calibrate the NN models. Figure~\ref{fig:s-basic-avg-slope} shows the calibration plots when we linearly shift the waveform samples and use NNs for time prediction. It can be seen that all three models (FC, CNN and LSTM) have successfully learned the linear dependency on input data without explicit labels in training. By carefully examining the figures, we find two interesting phenomenons: first, from the appearance of these figures, LSTM achieves the highest linearity when compared to the other two models; second, from the calculated average slope, CNN achieves the largest absolute value when compared to the other two models. Intuitively, LSTM might be the best of the three because it conforms better to the prior (linearly shifted predictions corresponding to linearly shifted input data); practically, CNN could also be used, because it makes the best use of training data to reduce the loss function (reflected by the average slope), although it seems to fluctuate in its predictions when shifting the input data linearly.

The well-trained NN models are evaluated on the test dataset. For each example composed of waveform samples coming from 8 analog channels, we use the model to predict the arrival time for individual channels. These predictions are fitted to a linear function indicated by the physical constraint. In the left column of figure~\ref{fig:nn-hist}, we show the distribution of Pearson correlation coefficient (R value), an indicator to judge the goodness of the linear fit, for all examples in the test dataset. It can be seen that, in spite of the noisy data and random interactions, NN models are able to reveal the linear correlation underlying the multi-channel input data. If we define examples with R value above 0.7 as ``good examples'', the ratios of good examples for FC, CNN and LSTM are 70.97\%, 77.93\% and 79.66\%. In the right column of figure~\ref{fig:nn-hist}, when calculating the time resolution with triples of adjoined analog channels on examples above the 0.7 threshold, the standard deviations for FC, CNN and LSTM are 476 \si{ps}, 384 \si{ps} and 424 \si{ps}. It is noteworthy that LSTM achieves the highest linearity and CNN achieves the best time resolution, which is in accord with the phenomenons in the calibration plots.

\begin{figure}[htb]
	\centering
	\includegraphics[width=0.75\textwidth]{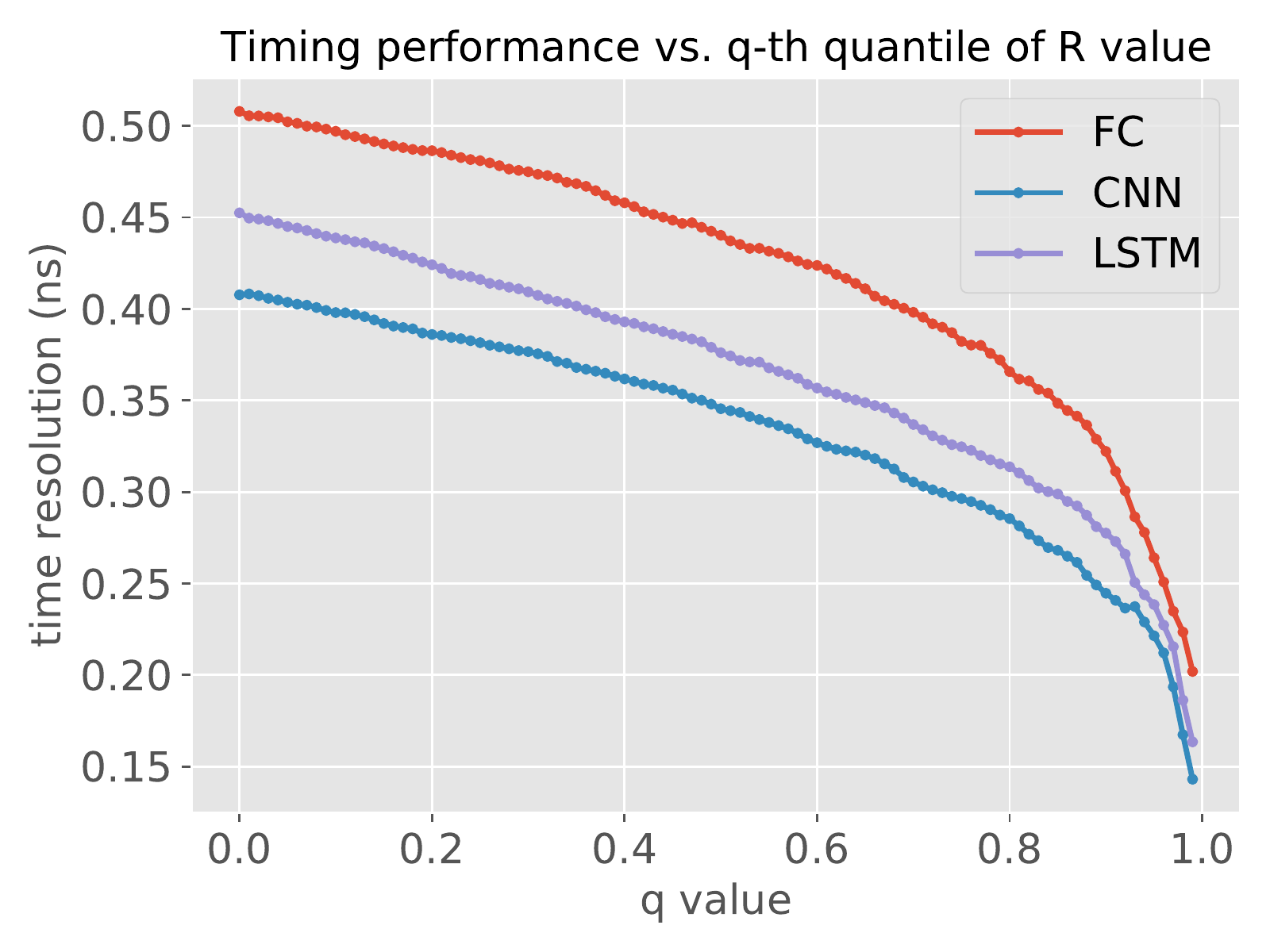}
	\caption{Comparison of timing performance of NN models at different q-th quantiles of the R value.}
	\label{fig:nn-s-basic-compare}
\end{figure}

In order to analyse the dependence of time resolution on R value, and to fairly compare NN models with equal numbers of examples in the test dataset, we draw the timing performance versus the q-th quantile of R value, as illustrated in figure~\ref{fig:nn-s-basic-compare}. When increasing the q value, decreasing number of examples with improved linearity are included for calculation. It can be seen that, in the aspect of time resolution on the same quantile, CNN performs steadily better than LSTM, and LSTM performs steadily better than FC. There is a nearly fixed gap between them when q value is below 0.8, and the gap quickly narrows when q value is 0.8 or larger.

\begin{figure}[htb]
	\centering
	\includegraphics[width=0.75\textwidth]{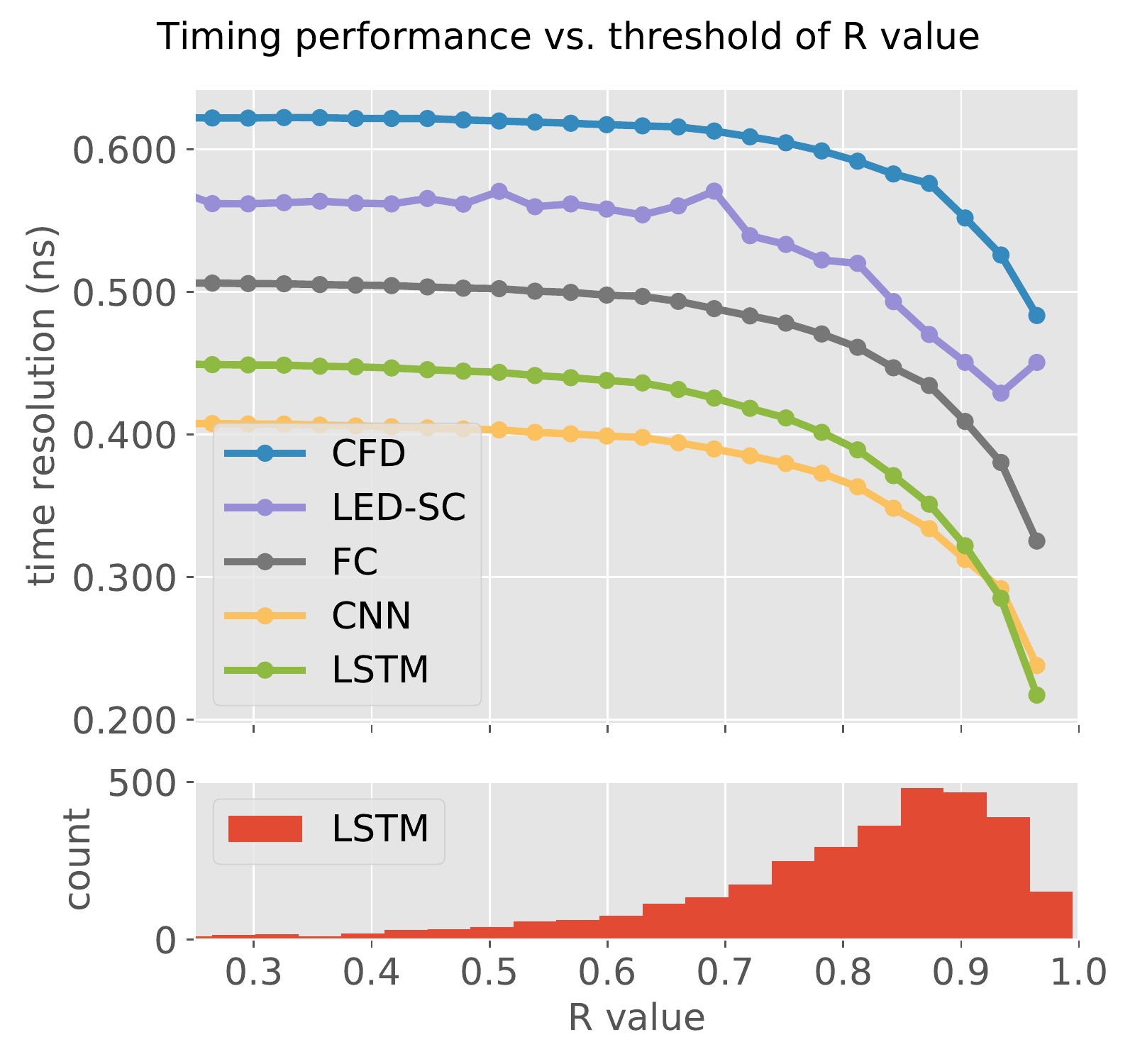}
	\caption{The lower histogram shows the distribution of the R value with the LSTM model. Based on the histogram, the upper plot shows the timing performance of two traditional methods (digital constant fraction discrimination and leading edge discrimination with slewing correction) and also NNs at different R value thresholds.}
	\label{fig:nn-tr-compare}
\end{figure}

Finally, we evaluate the NN models in comparison with other traditional timing methods. Two well-known methods, already validated in many real-world circumstances, are studied: the digital constant fraction discrimination (CFD) and the leading edge discrimination with slewing correction (LED-SC). The parameters used when applying these two methods are discussed in \ref{sec:detail-conf-ecal}. To make the comparison, we use the distribution of R value by LSTM as a reference to select examples with the particular linearity and to test all methods with the same examples. The test result is shown in figure~\ref{fig:nn-tr-compare}. It can be seen that, the performance of traditional methods is on the same order as NN models, but is apparently worse in this experimental condition. Since the ambient noise is significant and the signal-to-noise ratio is relatively low, accurate timing is challenging for traditional methods based on partial information in waveform samples, while NNs are more competent at this work by exploiting information in the input data as much as possible.

\subsubsection{Discussion}

Based on the observations above, we make the following points:

\begin{enumerate}
	\item In general, NN models work pretty well as desired when trained with the physical constraint in the label-free setting. Among them, LSTM achieves the highest linearity in its prediction, while CNN achieves the best performance. Which model is used in practice should be judged by the particular needs and also the objectives of measurement.
	\item In these two experiments, the physical quantity being learned is directly a time-of-arrival in the ADC clock domain, rather than the timing difference of physically coupled measurements \cite{Berg_2018,10038575}. This could be a prominent advantage if the detector works in stand-alone conditions, or in conditions where the matching property is not present. Besides, the self-supervised learning scheme eliminates the need of external equipment as timing reference \cite{GLADEN2020164505}.
	\item In the final part of the main results, we make a comparison between traditional methods and NNs and show the advantage of the latter. It should be mentioned that traditional methods can also reach the theoretically best resolution in proper conditions \cite{Ai_2021}, which can be satisfied by sophisticated design. However, NNs are promising in the long run, because they provide more flexibility and adaptability with guaranteed performance. More benefits are expected when hardware acceleration of NNs on front-end electronics is adopted in more extensive situations.
\end{enumerate}

\section{Conclusion}

In this paper, we propose a deep learning-based methodology and algorithmic implementation to analyse the timing performance of SiPM-based modularized detectors without explicitly labelling examples of event data. A formal mathematical framework is built to ensure the existence of the optimal mapping function desired by the method. The major innovation of the paper is to present a physics-constrained loss function and its associated regularizer customized for waveform sampling systems. Two experiments (toy experiment, NICA-MPD ECAL experiment) demonstrate the superior performance and adequate physical implication of NN models. In the toy experiment, timing resolution of less than 10 \si{ps} has been achieved with a dual-channel coincidence setup. In the ECAL experiment, a variety of NN models have been investigated and compared to traditional methods in terms of timing performance.

It should be noted that neural networks are versatile and able to be integrated with other feature extraction tasks, such as energy estimation \cite{Ai_2022} and position estimation \cite{Carra_2022}. Besides, the ability of deep learning to exploit information in noisy conditions indicates possible applications in harsh experimental environment, such as the time-of-flight measurement of laser-accelerated particles with interference of high-level electromagnetic pulses \cite{Salvadori2021}.

In the future, we will deploy the method on online processing hardware to achieve lower latency and higher energy efficiency. Existing \texttt{hls4ml} \cite{Aarrestad_2021,Khoda_2023} framework is proposed on FPGA platform with digital signal processors (DSP) available for fast inference (several \si{\micro\second} to several hundred \si{\micro\second} latency). On ASICs where no DSPs are ready for use, fully-customized neural network accelerators \cite{AI2020164420,10005128} are currently in development ($\sim$100 \si{\micro\second} latency).

\ack

This research is supported in part by the National Key Research and Development Program of China (under Grant Nos. 2020YFE0202001, 2022YFA1602103), in part by the National Natural Science Foundation of China (under Grant Nos. 12235006, 12075099, 11875146, U1932143), in part by the Fundamental Research Funds for the Central Universities (under Grant No. CCNU23XJ013) and in part by the China Postdoctoral Science Foundation (under Grant No. 2023M731244).

\appendix

\section{Proof of Proposition \ref{pro:main}}
\label{sec:proof}

\begin{proof}

Consider $f(x)$ is evaluated at $x_i = t_0 + a_i t_c + \Delta t_i + \Delta T_i$ where $i = 0, 1, ..., N-1$. Assume the argument of $L$ deviates from $f(x)$ by a small amount $\epsilon h(x)$, according to the calculus of variations:
\begin{IEEEeqnarray}{rCl}
	\frac{\delta L}{\delta f}(h) & := & \lim\limits_{\epsilon \to 0} \frac{L(f + \epsilon h) - L(f)}{\epsilon} \nonumber \\
	& = & \lim\limits_{\epsilon \to 0} \frac{1}{\epsilon} \int (I(\bm{Y} + \epsilon\bm{H}) - I(\bm{Y})) p(\bm{\Delta t}) p(\bm{\Delta T}) p(t_0) p(t_c) \mathrm{d}\bm{\Delta t} \mathrm{d}\bm{\Delta T} \mathrm{d}t_0 \mathrm{d}t_c \nonumber \\
	& = & 2 \int \bm{H}^T \bm{M} \bm{Y} p(\bm{\Delta t}) p(\bm{\Delta T}) p(t_0) p(t_c) \mathrm{d}\bm{\Delta t} \mathrm{d}\bm{\Delta T} \mathrm{d}t_0 \mathrm{d}t_c
\end{IEEEeqnarray}

\noindent where $\bm{H} = [ h(x_0), h(x_1), ..., h(x_{N-1}) ]^T$. If $f(x)$ is a linear function, i.e., $f(x) = kx + b$, we can simplify the above expression:
\begin{IEEEeqnarray}{rCl}
	\frac{\delta L}{\delta f}(h) & = & 2 \int \bm{H}^T \bm{M} ( (k-1)\bm{\Delta t} + k\bm{\Delta T} ) p(\bm{\Delta t}) p(\bm{\Delta T}) p(t_0) p(t_c) \mathrm{d}\bm{\Delta t} \mathrm{d}\bm{\Delta T} \mathrm{d}t_0 \mathrm{d}t_c \nonumber \\
	& = & 2 \int \bm{\breve{H}}^T \bm{M} ( (k-1)\bm{\Delta t} + k\bm{\Delta T} ) p(\bm{\Delta t}) p(\bm{\Delta T}) \mathrm{d}\bm{\Delta t} \mathrm{d}\bm{\Delta T} \label{equ:proof-linear-general}
\end{IEEEeqnarray}

\noindent where $\bm{\breve{H}} = [ \breve{h}_0(\Delta t_0 + \Delta T_0), \breve{h}_1(\Delta t_1 + \Delta T_1), ..., \breve{h}_{N-1}(\Delta t_{N-1} + \Delta T_{N-1}) ]^T$ which is a vector of deviation functions after $(t_0, t_c)$ are marginalized. Without loss of generality, we assume $\bm{\breve{H}} = [\delta(\Delta t_0 + \Delta T_0 - \tau), 0, ..., 0 ]^T$ as a ``probe'' vector for arbitrary functions. Substitute into equation (\ref{equ:proof-linear-general}), and consider the i.i.d condition and zero mean values for $\Delta t_i$ and $\Delta T_i$:
\begin{IEEEeqnarray}{rCl}
	\frac{\delta L}{\delta f}(h) & = & 2 M_{00} \int \delta(\Delta t_0 + \Delta T_0 - \tau) ( (k-1)\Delta t_0 + k\Delta T_0 ) p(\Delta t_0) p(\Delta T_0) \mathrm{d}\Delta t_0 \mathrm{d}\Delta T_0 \quad\quad \label{equ:proof-uni-linear}
\end{IEEEeqnarray}

\noindent where $M_{00}$ is the top-left element of the matrix $\bm{M}$. To compute equation (\ref{equ:proof-uni-linear}), we first use the method of substitution to normalize $p(\Delta t_0), p(\Delta T_0)$ to the standard normal density  $\tilde{p}(\cdot) \sim \mathcal{N}(0, 1)$, and then rotate the axis with central symmetry to align with the delta function:
\begin{IEEEeqnarray}{rCl}
	\frac{\delta L}{\delta f}(h) & = & 2 M_{00} \int \delta(\sigma_1 x + \sigma_2 y - \tau) ( (k-1) \sigma_1 x + k \sigma_2 y ) \tilde{p}(x) \tilde{p}(y) \mathrm{d}x \mathrm{d}y \nonumber \\
	& = & 2 M_{00} \int \delta(\sqrt{\sigma_1^2 + \sigma_2^2}u - \tau) \frac{ ( (k-1)\sigma_1^2 + k\sigma_2^2 ) u + \sigma_1\sigma_2 v }{\sqrt{\sigma_1^2 + \sigma_2^2}} \tilde{p}(u) \tilde{p}(v) \mathrm{d}u \mathrm{d}v \nonumber \\
	& = & 2 M_{00} \left. \cdot \frac{ ( (k-1)\sigma_1^2 + k\sigma_2^2 ) u}{\sigma_1^2 + \sigma_2^2} \tilde{p}(u) \right|_{u= \tau / \sqrt{\sigma_1^2 + \sigma_2^2 }} \nonumber \\
	& = & 2 M_{00} \cdot \frac{ ( (k-1)\sigma_1^2 + k\sigma_2^2 ) \tau}{(\sigma_1^2 + \sigma_2^2)^{3/2}} \cdot \tilde{p}(\frac{ \tau }{ \sqrt{\sigma_1^2 + \sigma_2^2 }})
\end{IEEEeqnarray}

By setting the above expression to zero, we can solve for $k = \frac{\sigma_1^2}{\sigma_1^2 + \sigma_2^2}$. Similarly, for other ``probe'' vectors (such as $[0, ..., 0, \delta(\Delta t_i + \Delta T_i - \tau), 0, ..., 0]$), we can get the same conclusion. It can be generalized to arbitrary deviation functions.

By substituting $f(x) = kx + b$ into the expression of $L(f)$, we can get the minimum value:
\begin{IEEEeqnarray}{rCl}
	L_{\mathrm{min}} & = & \int ((k-1)\bm{\Delta t} + k\bm{\Delta T})^T \bm{M} ((k-1)\bm{\Delta t} + k\bm{\Delta T}) p(\bm{\Delta t}) p(\bm{\Delta T}) \mathrm{d}\bm{\Delta t} \mathrm{d}\bm{\Delta T} \nonumber \\
	& = & \int \sum\limits_{i=0}^{N-1}M_{ii} ((k-1)^2 \Delta t_i^2 + k^2 \Delta T_i^2) p(\bm{\Delta t}) p(\bm{\Delta T}) \mathrm{d}\bm{\Delta t} \mathrm{d}\bm{\Delta T} \nonumber \\
	& = & \sum\limits_{i=0}^{N-1}M_{ii} ((k-1)^2 \sigma_1^2 + k^2 \sigma_2^2) \nonumber \\
	& = & \frac{\sigma_1^2 \sigma_2^2}{\sigma_1^2 + \sigma_2^2} \mathrm{\textbf{tr}}(\bm{M})
\end{IEEEeqnarray}

Now we have proved the sufficient condition in Proposition \ref{pro:main}.

\end{proof}

\section{Details of neural network architectures}
\label{sec:detail-nn-arch}

\subsection{Toy experiment}
\label{sec:detail-nn-arch-toy}

The CNN model used in the toy experiment is shown in table~\ref{tab:nn-arch-toy}. For convolution layers, the kernel width is fixed at 4, and the stride is fixed at 2.

\begin{table}[H]
	\centering
	\caption{\label{tab:nn-arch-toy} The architecture of CNN in the toy experiment.}
	\begin{tabularx}{\textwidth}{XXXXXXX}
		\hline
		& Conv\_1 & Conv\_2 & Conv\_3 & FC\_1 & FC\_2 & FC\_3 \\
		\hline
		in shape & 2048$\times$1 & 1024$\times$8 & 512$\times$16 & 8192 & 64 & 64 \\
		out shape & 1024$\times$8 & 512$\times$16 & 256$\times$32 & 64 & 64 & 1 \\
		kernel & 4$\times$1$\times$8 & 4$\times$8$\times$16 & 4$\times$16$\times$32 & 8192$\times$64 & 64$\times$64 & 64$\times$1 \\
		bias & 8 & 16 & 32 & 64 & 64 & 1 \\
		activation & ReLU & ReLU & ReLU & ReLU & ReLU & Linear \\    
		\hline
	\end{tabularx}
\end{table}

\subsection{NICA-MPD ECAL experiment}
\label{sec:detail-nn-arch-ecal}

The FC, CNN and LSTM models used in the ECAL experiment are shown in table~\ref{tab:nn-arch-ecal-fc}, table~\ref{tab:nn-arch-ecal-cnn} and table~\ref{tab:nn-arch-ecal-lstm}, respectively. For convolution layers, the kernel width is fixed at 4, and the stride is fixed at 2. For LSTM layers, we use the LSTM class in the Keras package and make each unit generate a sequence instead of a single value.

\begin{table}[H]
	\centering
	\caption{\label{tab:nn-arch-ecal-fc} The architecture of FC in the ECAL experiment.}
	\begin{tabularx}{\textwidth}{XXXXXX}
		\hline
		& FC\_1 & FC\_2 & FC\_3 & FC\_4 & FC\_5 \\
		\hline
		in shape & 800 & 64 & 64 & 64 & 64 \\
		out shape & 64 & 64 & 64 & 64 & 1 \\
		kernel & 800$\times$64 & 64$\times$64 & 64$\times$64 & 64$\times$64 & 64$\times$1 \\
		bias & 64 & 64 & 64 & 64 & 1 \\
		activation & ReLU & ReLU & ReLU & ReLU & Linear \\    
		\hline
	\end{tabularx}
\end{table}

\begin{table}[H]
	\centering
	\caption{\label{tab:nn-arch-ecal-cnn} The architecture of CNN in the ECAL experiment.}
	\begin{tabularx}{\textwidth}{XXXXXXX}
		\hline
		& Conv\_1 & Conv\_2 & Conv\_3 & FC\_1 & FC\_2 & FC\_3 \\
		\hline
		in shape & 800$\times$1 & 400$\times$8 & 200$\times$16 & 3200 & 64 & 64 \\
		out shape & 400$\times$8 & 200$\times$16 & 100$\times$32 & 64 & 64 & 1 \\
		kernel & 4$\times$1$\times$8 & 4$\times$8$\times$16 & 4$\times$16$\times$32 & 3200$\times$64 & 64$\times$64 & 64$\times$1 \\
		bias & 8 & 16 & 32 & 64 & 64 & 1 \\
		activation & ReLU & ReLU & ReLU & ReLU & ReLU & Linear \\    
		\hline
	\end{tabularx}
\end{table}

\begin{table}[H]
	\centering
	\caption{\label{tab:nn-arch-ecal-lstm} The architecture of LSTM in the ECAL experiment.}
	\begin{tabularx}{\textwidth}{XXXXXXX}
		\hline
		& LSTM\_1 & LSTM\_2 & LSTM\_3 & FC\_1 & FC\_2 & FC\_3 \\
		\hline
		in shape & 800$\times$1 & 800$\times$20 & 800$\times$10 & 4000 & 64 & 64 \\
		out shape & 800$\times$20 & 800$\times$10 & 800$\times$5 & 64 & 64 & 1 \\
		kernel & 1$\times$80 & 20$\times$40 & 10$\times$20 & 4000$\times$64 & 64$\times$64 & 64$\times$1 \\
		recurrent & 20$\times$80 & 10$\times$40 & 5$\times$20 & -- & -- & -- \\
		bias & 80 & 40 & 20 & 64 & 64 & 1 \\
		activation & tanh & tanh & tanh & ReLU & ReLU & Linear \\    
		\hline
	\end{tabularx}
\end{table}

\section{Details of the configurations}
\label{sec:detail-conf}

\begin{figure}[htb]
	\centering
	\begin{subfigure}[b]{0.45\textwidth}
		\centering
		\includegraphics[width=\textwidth]{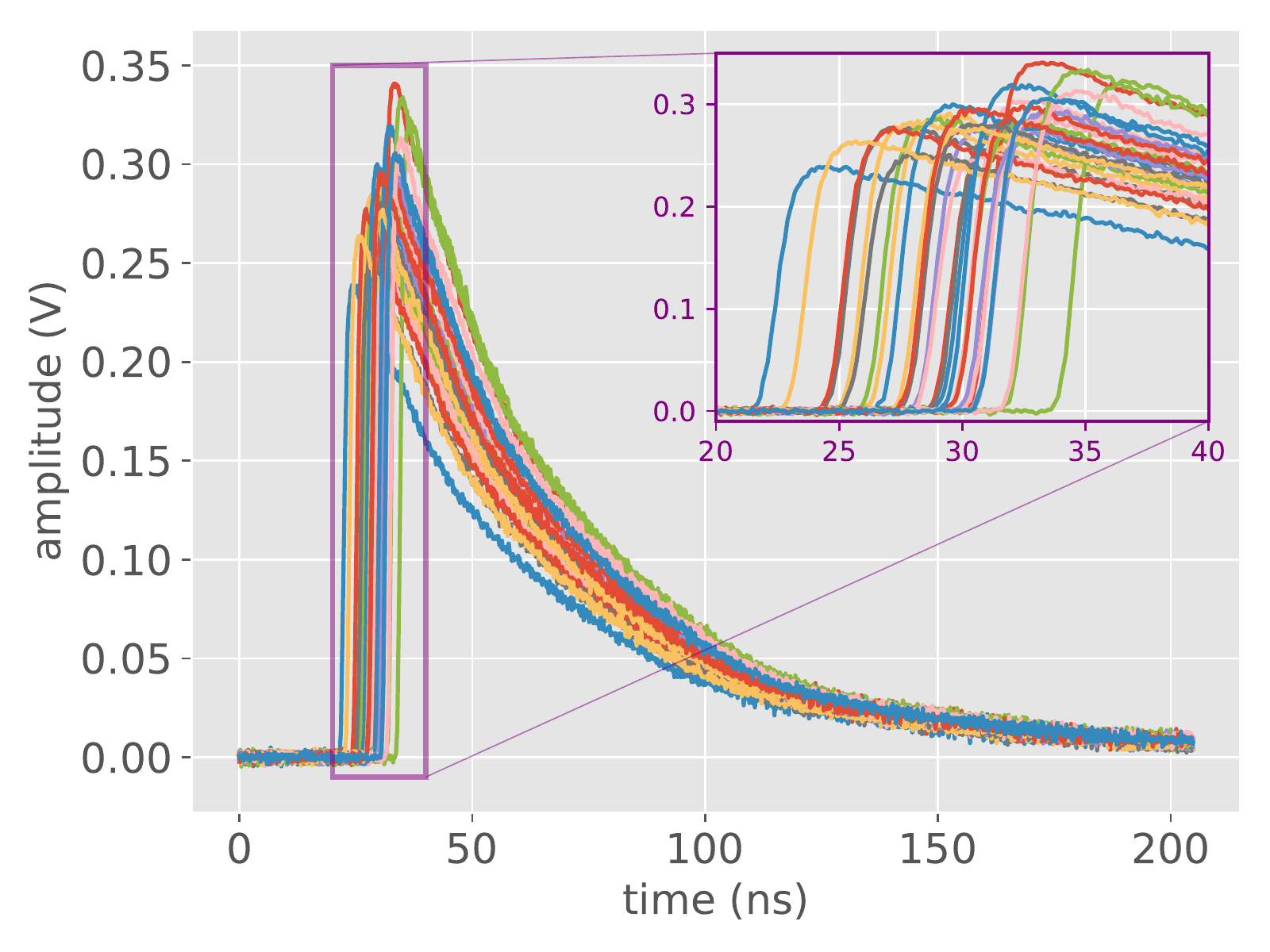}
		\caption{Toy experiment}
		\label{fig:s-toy-data-inspect}
	\end{subfigure}
	\begin{subfigure}[b]{0.45\textwidth}
		\centering
		\includegraphics[width=\textwidth]{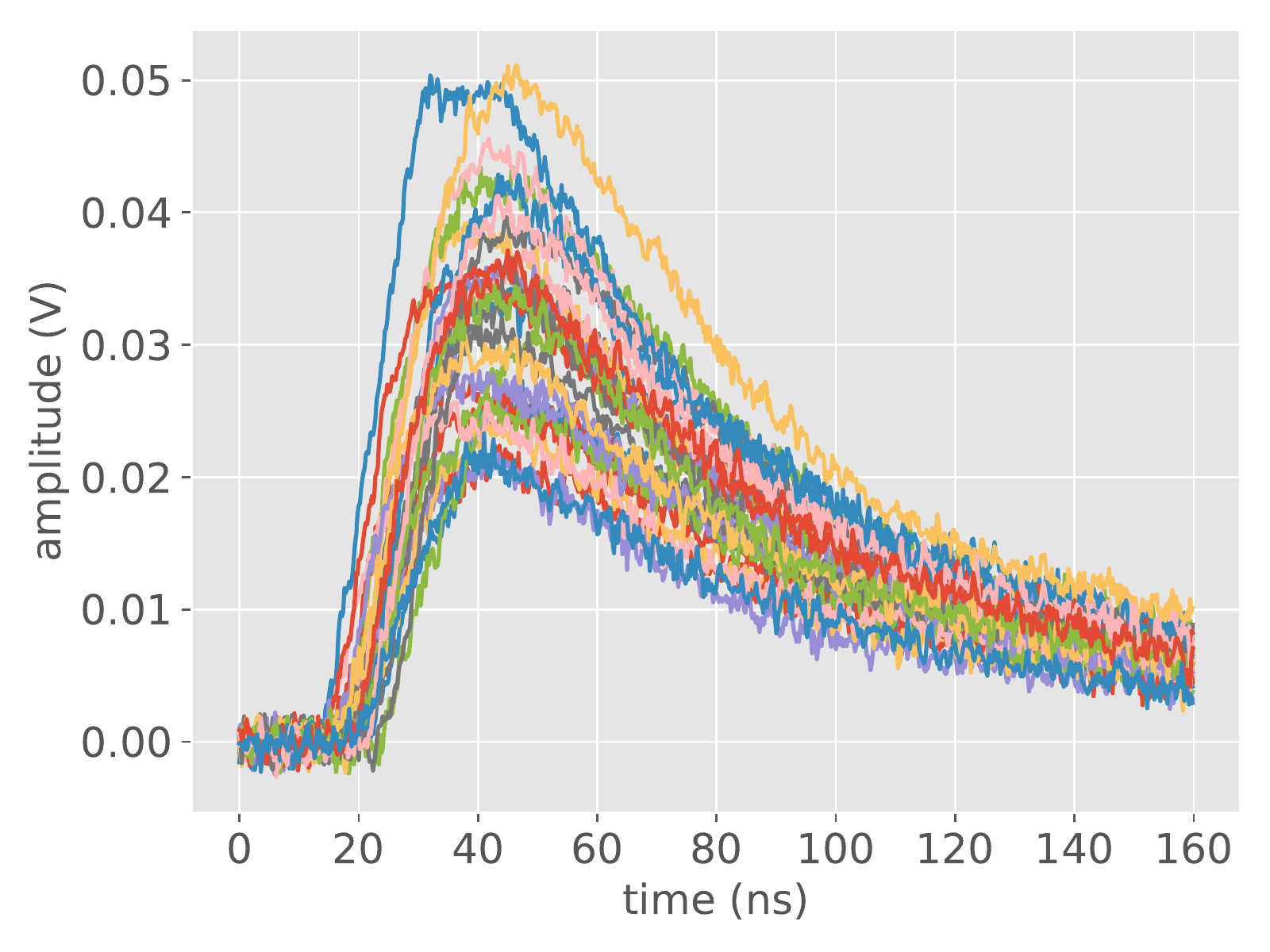}
		\caption{ECAL experiment}
		\label{fig:s-basic-data-inspect}
	\end{subfigure}
	\caption{Visualization of single-channel waveform samples in (a) the toy experiment and (b) the NICA-MPD ECAL experiment.}
	\label{fig:data-inspect}
\end{figure}

\begin{figure}[htb]
	\centering
	\begin{subfigure}[b]{0.45\textwidth}
		\centering
		\includegraphics[width=\textwidth]{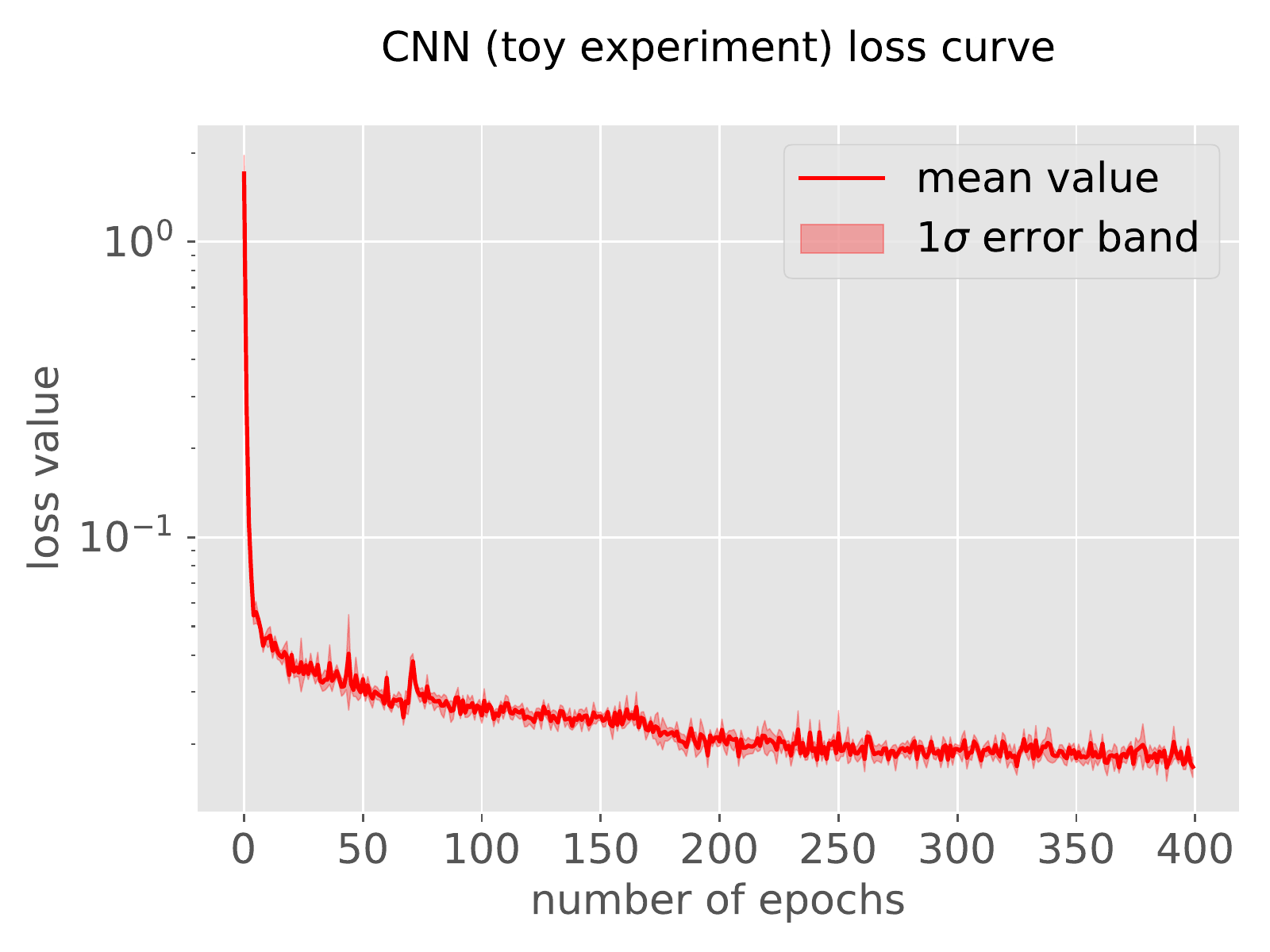}
		\caption{Toy experiment}
		\label{fig:s-toy-loss-curve}
	\end{subfigure}
	\begin{subfigure}[b]{0.45\textwidth}
		\centering
		\includegraphics[width=\textwidth]{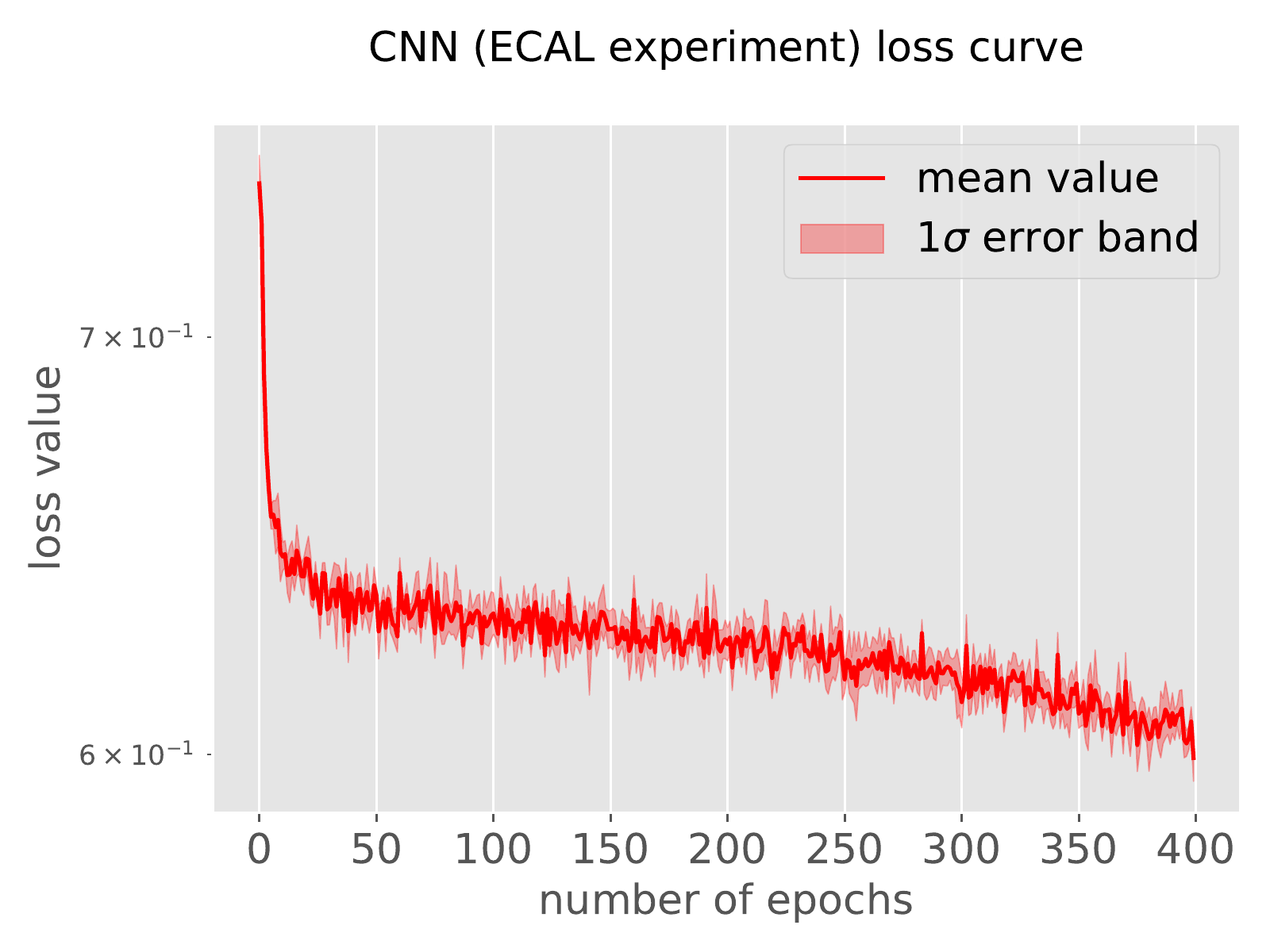}
		\caption{ECAL experiment}
		\label{fig:cnn-s-basic-loss-curve}
	\end{subfigure}
	\caption{Visualization of training losses of CNNs in (a) the toy experiment and (b) the NICA-MPD ECAL experiment.}
	\label{fig:loss-curve}
\end{figure}

In experiments below, we perform neural network training on a laptop computer with a NVIDIA GeForce RTX 2070 graphic card (8 GB video memory). The training lasts for several minutes to one hour depending on the network complexity.

\subsection{Toy experiment}
\label{sec:detail-conf-toy}

In the toy experiment, 10024 examples of paired waveform samples are collected in total. The data collection takes 17 minutes at a rate of about 10 examples per second. The amplitudes of these examples are normalized by linearly scaling and shifting to the range [0.05, 0.95]. Then these examples are divided into the training dataset and the test dataset with the ratio 4:1.

In training, for each example, a series of 2048 waveform samples (204.8-\si{ns} time window) is selected from 4000 original samples. Figure \ref{fig:s-toy-data-inspect} illustrates 30 such examples. To improve the variations of data, we randomly choose the origin (starting point) of the time window in a range of 50 sampling intervals (5 \si{ns}). The batch size we use is 32, and the training lasts for 400 epochs. The training loss curve is shown in figure \ref{fig:s-toy-loss-curve}. We calibrate the base model with the training dataset, and evaluate on all examples in the test dataset.

For comparison, digital CFD is applied to the same examples in the test dataset. After the (optional) low-pass filter, the maximum of the waveform is recorded as $S_{\mathrm{max}}$, and the time point crossing the $0.5 \cdot S_{\mathrm{max}}$ (with linear interpolation between nearby samples) is regarded as the prediction.

\subsection{NICA-MPD ECAL experiment}
\label{sec:detail-conf-ecal}

In the ECAL experiment, 16178 examples of grouped waveform samples are collected in total. The data collection takes 18 days at a rate of about 38 examples per hour. The normalization and dataset division (4:1 ratio) are similar to the toy experiment.

In training, for each example, a series of 800 waveform samples (160-\si{ns} time window) is selected from 1000 original samples. Figure \ref{fig:s-basic-data-inspect} illustrates 30 such examples. To improve the variations of data, we randomly choose the origin (starting point) of the time window in a range of 25 sampling intervals (5 \si{ns}). The batch size we use is 32. For FC and CNN, the training lasts for 400 epochs, and 50 epochs for LSTM. For CNN, the training loss curve is shown in figure \ref{fig:cnn-s-basic-loss-curve}. Each base model is calibrated with the training dataset and evaluated on the test dataset.

To test digital CFD, a second-order low-pass filter with 0.02-\si{GHz} critical frequency (at 5-\si{GHz} sampling rate) applies to waveform samples before the timing method (the critical frequency is picked to produce the best results). The threshold is also set at $0.5 \cdot S_{\mathrm{max}}$.

To test LED-SC, the leading edge of the waveform is fitted to a third-order polynomial function before solving for the time point crossing a fixed threshold. The time residual is calculated with the triple of adjoined analog channels, and the amplitudes are recorded alongside the time. To correct the timing result, the following time-amplitude relation is used:

\begin{equation}
	T_{\mathrm{residual}} = p_0 + \frac{p_1}{\sqrt{A_i}} + \frac{p_2}{A_i} + p_3 \cdot A_i
\end{equation}

\noindent where $A_i$ is the amplitude from one of the analog channels. We fit all examples to the above function with the amplitude from each channel by turns. We correct the time residuals by subtracting the fitting results depending only on the amplitude. The correction is repeated 3 rounds.

\section*{References}

\bibliography{mybibfile}

\end{document}